\documentclass{LMCS}

\def\dOi{9(4:13)2013}
\lmcsheading%
{\dOi}
{1--26}
{}
{}
{Nov.~14, 2012}
{Nov.~14, 2013}
{}

\ACMCCS{[{\bf Theory of computation}]: Models of computation---Timed
  and hybrid models; Models of computation---Concurrency---Distributed
  computing models}

\usepackage{hyperref}
\usepackage{mathtools}
\usepackage{wasysym}
\usepackage[norelsize,boxed,lined,longend]{algorithm2e}
\usepackage[tight,footnotesize]{subfigure}
\usepackage{tikz}
\usetikzlibrary{automata,shapes,arrows,decorations,petri}
\usetikzlibrary{positioning}
\usetikzlibrary{calc}
\tikzstyle{transition} = [rectangle,
                   draw,
                   inner sep      = 0,
                   text width     = 0.4cm,
                   minimum height = 0.4cm,
                   text centered]
\tikzstyle{state} = [circle,
                   draw,
                   inner sep      = 0,
                   text width     = 0.4cm,
                   minimum height = 0.4cm,
                   text centered]
\tikzstyle{none} = [draw = none]
\tikzstyle{sad} = [none,circle,inner sep=-2.5pt]
\tikzstyle{every picture} = [auto,
                   >             = stealth,
                   node distance = 1.2cm,
                   initial text=,
                   font=\footnotesize,
                   initial distance=0.3cm]
\usepackage{macros}

\theoremstyle{plain}

\def\ie{{\em i.e.\ }}

\begin{document}

\title[Avoiding Shared Clocks in NTA]
{Avoiding Shared Clocks in Networks of Timed Automata\rsuper*}

\author[S.~Balaguer]{Sandie Balaguer\rsuper a}	
\address{LSV (ENS Cachan, CNRS, Inria)\\
61, avenue du Pr\'esident Wilson\\
94235 CACHAN Cedex, France}	
\email{\{balaguer,chatain\}@lsv.ens-cachan.fr}  
\thanks{{\lsuper a}This work is partially supported by the French ANR project ImpRo.}

\author[T.~Chatain]{Thomas Chatain\rsuper b}	


\titlecomment{{\lsuper*}This paper extends the version presented at
CONCUR'12~\cite{BC-concur12}.}


\begin{abstract}
  \noindent Networks of timed automata (NTA) are widely used to model distributed real-time
  systems. Quite often in the literature, the automata are allowed to share
  clocks, \ie transitions of one automaton may be guarded by conditions
  on the value of clocks reset by another automaton. This is a
  problem when one considers implementing such model in a distributed
  architecture, since reading clocks a priori requires communications which are
  not explicitly described in the model. We focus on the following question: given
  an NTA $A_1\parallel A_2$ where $A_2$ reads some clocks reset by $A_1$,
  does there exist an NTA $A'_1\parallel A'_2$ without shared clocks
  with the same behavior as the initial NTA?
  For this, we allow the automata to exchange information during synchronizations
  only, in particular by copying the value of their neighbor's clocks.
  We discuss a formalization of the problem and define an appropriate
  behavioural equivalence. Then we give a criterion using the notion
  of contextual timed transition system, which represents the behavior of $A_2$
  when in parallel with $A_1$.
  Finally, we effectively build $A'_1\parallel A'_2$ when it exists.
\end{abstract}

\maketitle

\section*{Introduction}
Timed automata \cite{AD94} are one of the most famous formal models for
real-time systems. They have been deeply studied and very mature tools are
available, like \textsc{Uppaal} \cite{uppaal}, \textsc{Epsilon}
\cite{Epsilon93} and \textsc{Kronos} \cite{kronos}.

Networks of Timed Automata (NTA) are a natural generalization to model real-time
distributed systems. In this formalism, each automaton has a set of clocks that
constrain its real-time behavior. But quite often in the literature, the
automata are allowed to share clocks, which provides a special way of making the
behavior of one automaton depend on what the others do. Actually shared clocks
are relatively well accepted and can be a convenient feature for modeling
systems. Imagine for instance several agents performing together a distributed
task according to a predefined schedule. In a typical implementation the
schedule would be sent to the agents at the beginning and every agent would
store its own copy of the schedule. But for a (simplified) model of the system,
it may be easier to have one timed automaton modeling a single copy of the
schedule and every agent referring to it via shared clocks.

Since NTA are almost always given a sequential semantics,
shared clocks can be handled very easily even by tools: once the NTA is
transformed into a single timed automaton by the classical product construction,
the notion of distribution is lost and the notion of shared clock itself becomes
meaningless. Nevertheless, implementing a model with shared clocks in a
multi-core architecture is not straightforward since reading clocks a priori
requires communications which are not explicitly described in the model.

Here we are concerned with the expressive power of shared clocks according to
the distributed nature of the system. We are aware of only one previous study
about this aspect, presented in~\cite{LPW07}.
Our purpose is to identify NTA where sharing clocks could be avoided,  \ie NTA
which syntactically use shared clocks, but whose semantics could be achieved by
another NTA without shared clocks. For simplicity, we look at NTA made of two
automata $A_1$ and $A_2$ where only $A_2$ reads clocks reset by $A_1$.
The first step is to formalize which aspect of the semantics we want to preserve
in this setting. Then the idea is essentially to detect cases where $A_2$ can
avoid reading a clock because its value does not depend on the actions that are
local to $A_1$ and thus unobservable to $A_2$. To generalize this idea we
have to compute the knowledge of $A_2$ about the state of $A_1$. We show
that this knowledge is maximized if we allow $A_1$ to communicate its state to
$A_2$ each time they synchronize on a common action.

In order to formalize our problem we need an appropriate notion of behavioral
equivalence between two NTA\@. We explain why classical comparisons based on the
sequential semantics, like timed bisimulation, are not sufficient here.
We need a notion that takes the distributed nature of the system into account.
That is, a component cannot observe the moves and the state of the other and
must choose its local actions according to its partial knowledge of the state of
the system. We define the notion of contextual timed transition systems
(contextual TTS) in order to formalize this idea.

Then we express the problem of avoiding shared clocks in terms of contextual TTS
and we give a characterization of the NTA for which shared clocks can be avoided.
Finally we effectively construct an NTA without shared clocks with the same
behavior as the initial one, when it exists.
A possible interest is to allow a designer to use shared clocks as a high-level
feature in a model of a protocol, and rely on our transformation to make it
implementable.

\subsubsection*{Related work}

The semantics of time in distributed systems has already been debated. The idea
of localizing clocks has already been proposed and some authors
\cite{ABGMN-concur08,Dima,Bengtsson98partialorder} have even suggested to use
local-time semantics with independently evolving clocks. Here we stay in the
classical setting of perfect clocks evolving at the same speed. This is a key
assumption that provides an implicit synchronization and lets us know some clock
values without reading them.

Many formalisms exist for real-time distributed systems, among which
NTA~\cite{AD94} and time Petri nets~\cite{Merlin}. So far, their
expressiveness was
compared~\cite{BCHRL-FORMATS2005,BoyerR08,Cassez-JSS06,Srba08} essentially in terms of
sequential semantics that forget concurrency. In \cite{BCH-fmsd12}, we defined a
concurrency-preserving translation from time Petri nets to networks of timed
automata. This transformation uses shared clocks and the question whether these
could be avoided remained open.

While partial-order semantics and unfoldings are well known for untimed systems,
they have been very little studied for distributed real-time
systems~\cite{Cassez-Ch-Jard_ATVA06,BHR-atva06}. Partial order reductions for (N)TA were
proposed in~\cite{Minea99,Bengtsson98partialorder,LugiezNZ05}.
Behavioral equivalence relations for distributed systems, like
history-preserving bisimulations, were defined for
untimed systems only \cite{BestDKP91,GlabbeekG01}.

Finally, our notion of contextual TTS deals with knowledge of agents in
distributed systems. This is the aim of epistemic logics
\cite{ReasoningAboutKnowledge},
which have been extended to real-time in \cite{WoznaL04, Dima09}.
Our notion of contextual TTS also resembles the technique of partitioning
states based on observation, used in timed games with partial observability~\cite{Bouyer03,David09}.

\subsubsection*{Organization of the paper}
The paper is organized as follows. Section~\ref{sec:preliminaries} recalls basic
notions about TTS and NTA\@. Section~\ref{sec:nsc_problem_setting} presents the
problem of avoiding shared clocks on examples and rises the problem of comparing
NTA component by component. For this, the notion of contextual TTS is developed
in Section~\ref{sec:TTS}. The problem of avoiding shared clocks is formalized
and characterized in terms of contextual TTS\@. Then
Section~\ref{sec:construction} presents our construction.


\section{Preliminaries}\label{sec:preliminaries}                               %
\subsection{Timed Transition Systems}
The behavior of timed systems is often described as timed transition systems.
\begin{defi}
  A \emph{timed transition system} (TTS)
  is a tuple \mbox{$(S, s_0 , \Sigma, {\rightarrow})$}
  where
  $S$ is a set of states,
  $s_0 \in S$ is the initial state,
  $\Sigma$ is a finite set of actions disjoint from $\Reals$, and
  ${\rightarrow} \subseteq S \times (\Sigma \cup \Reals ) \times S$
  is a set of edges.
\end{defi}
For any $a\in\Sigma\cup\Reals $, we write $s\xrightarrow{a}s'$ if
\mbox{$(s,a,s')\in{\rightarrow}$}, and $s\xrightarrow{a}$ if for some $s'$,
\mbox{$(s,a,s')\in{\rightarrow}$}. 
%
We define the transition relation $\Rightarrow$ as:
\begin{itemize}
  \item $s\xRightarrow{\eps} s'$ if $s(\xrightarrow{\eps})^*s'$,
  \item $\forall a\in\Sigma$,
  $s\xRightarrow{a} s'$ if $s(\xrightarrow{\eps})^*
  \xrightarrow{a}(\xrightarrow{\eps})^* s'$,
  \item $\forall d\in\Reals$,
  $s\xRightarrow{d} s'$ if $s(\xrightarrow{\eps})^*
  \xRightarrow{d_0}(\xrightarrow{\eps})^*\cdots
  \xRightarrow{d_n}(\xrightarrow{\eps})^* s'$, where
  $\sum_{k=0}^n d_k=d$.
\end{itemize}

A \emph{path} of a TTS is a possibly infinite sequence of transitions
$\rho = s\xrightarrow{d_0}s_0'\xrightarrow{a_0}\cdots
s_n\xrightarrow{d_n}s_n'\xrightarrow{a_n}\cdots$, where, for all $i$,
$d_i\in\Reals$ and $a_i\in\Sigma$.
A path is \emph{initial} if it starts in $s_0$.
A path $\rho = s\xrightarrow{d_0}s_0'\xrightarrow{a_0}\cdots
s_n\xrightarrow{d_n}s_n'\xrightarrow{a_n}s'_n\cdots$ generates a
\emph{timed word} \mbox{$w=(a_0,t_0)(a_1,t_1)\dots(a_n,t_n)\dots$} where, for
all $i$, $t_i=\sum_{k=0}^i d_k$. The duration of $w$ is
$\delta(w)=\sup_i t_i$ and the untimed word of $w$ is
$\lambda(w)=a_0 a_1 \dots a_n\dots$. 
$\TW{\Sigma}$ denotes the set of finite timed words of duration $0$
over $\Sigma$, i.e.\ $\TW{\Sigma}=\{w\mid\delta(w)=0\land\lambda(w)\in\Sigma^*\}$. $\paths{\Sigma,d}$ denotes the set of finite paths of duration $d$
over $\Sigma$.
Lastly, we write $s\xrightarrow{w}s'$ if there is a path from $s$ to $s'$
that generates the timed word $w$.

In the sequel, 
we use the following notations: for $i\in\{1,2\}$,
$T_i = (S_i, s_i^0, \Sigma_i,{\rightarrow_i})$ is a TTS, and
$\Sigma_{i}^{\not\eps}=\Sigma_i\setminus\{\eps\}$, where $\eps$
is the silent action.

\subsubsection*{Product of timed transitions systems.}
The \emph{product} of $T_1$ and $T_2$, denoted by $T_1\x T_2$, is the TTS
\mbox{$\left(S_1\times S_2, (s_1^0, s_2^0), \Sigma_1\cup\Sigma_2, {\rightarrow}\right)$},
where ${\rightarrow}$ is defined as:
\begin{itemize}
  \item $(s_1,s_2)\xrightarrow{a}(s_1',s_2)$ iff $s_1\xrightarrow{a}_1 s'_1$,
  for any $a\in\Sigma_1\setminus\Sigma_{2}^{\not\eps}$,
  \item $(s_1,s_2)\xrightarrow{a}(s_1,s_2')$ iff $s_2\xrightarrow{a}_2 s'_2$,
  for any $a\in\Sigma_2\setminus\Sigma_{1}^{\not\eps}$,
  \item $(s_1,s_2)\xrightarrow{a}(s_1',s_2')$ iff $s_1\xrightarrow{a}_1 s'_1$
  and $s_2\xrightarrow{a}_2 s'_2$, for any $a\in(\Sigma_{1}^{\not\eps}\cap\Sigma_{2}^{\not\eps})\cup\Reals$.
\end{itemize}

\subsubsection*{Timed Bisimulations}
Let ${\R}$ be a binary relation over $S_1 \times S_2$.
%
${\R}$ is a
\emph{strong (resp.\ weak) timed bisimulation} relation between $T_1$ and $T_2$ if
$s_1^0\R s_2^0$ and $s_1\R s_2$ implies that, for any
$a \in \Sigma\cup \Reals$,
  if $s_1\xrightarrow{a}_1 s'_1$, then, for some $s'_2$,
  $s_2\xrightarrow{a}_2 s'_2$ (resp.\ $s_2\xRightarrow{a}_2 s'_2$) and
  $s'_1\R s'_2$;
  and conversely, if $s_2\xrightarrow{a}_2 s'_2$, then, for some $s'_1$,
  $s_1\xrightarrow{a}_1 s'_1$ (resp.\ $s_1\xRightarrow{a}_1 s'_1$) and $s'_1\R s'_2$.

We write $T_1\stb T_2$ (resp.\ $T_1\wtb T_2$) when there is a strong (resp.\ weak)
timed bisimulation between $T_1$ and $T_2$.

\subsection{Networks of Timed Automata}
The set $\mathcal{B}(X)$ of clock constraints over the set of clocks $X$ is
defined by the grammar $g::=x\bowtie k \mid g \land g$, where $x \in X$,
$k \in \N$ and ${\bowtie} \in \{<, \leq, =, \geq, >\}$. Invariants are clock
constraints of the form $i::=x\leq k \mid x < k \mid i \land i$.
\begin{defi} 
  A \emph{network of timed automata (NTA)} \cite{AD94} is a parallel composition
  of timed automata (TA) denoted as \mbox{$A_1\parallel\cdots\parallel A_n$},
  with $A_i=(L_i,\ell_i^0,X_i,\Sigma_i,E_i,\inv_i)$ where
  $L_i$ is a finite set of \emph{locations},
  $\ell_i^0 \in L_i$ is the \emph{initial} location,
  $X_i$ is a finite set of \emph{clocks},
  $\Sigma_i$ is a finite set of \emph{actions},
  $E_i \subseteq L_i \times \mathcal{B}(X_i) \times \Sigma_i \times 2^{X_i}\times L_i$
  is a set of \emph{edges},
  and $\inv_i: L_i \rightarrow \mathcal{B}(X_i)$ assigns \emph{invariants}
  to locations.
\end{defi}
If $(\ell,g,a,r,\ell') \in E_i$, we also write $\ell\xrightarrow{g,a,r}\ell'$.
For such an edge, 
$g$ is the \emph{guard}, $a$ the \emph{action} 
and $r$ the set of clocks to \emph{reset}.
$C_i\subseteq X_i$ is the set
of clocks reset by $A_i$ and for $i\neq j$, $C_i\cap C_j$ may not be
empty.

\subsubsection*{Semantics} For simplicity, we give the semantics of a network of two TA
$A_1\parallel A_2$. We denote by $((\ell_1,\ell_2), v)$ a \emph{state} of
the NTA, where  $\ell_1$ and $\ell_2$ are the current locations, and
$v:X\rightarrow \Reals$, with $X=X_1\cup X_2$, is a \emph{clock valuation}
that maps each clock to its current value. A state is legal only if its
valuation $v$ satisfies the invariants of the current locations, denoted by
$v \models \inv_1(\ell_1)\land \inv_2(\ell_2)$.
For each set of clocks $r\subseteq X$, the valuation $v[r]$ is defined by
$v[r](x)=0$ if $x\in r$ and $v[r](x)=v(x)$ otherwise.
For each $d \in \Reals$, the valuation $v+d$ is
defined by $(v+d)(x)=v(x)+d$ for each $x \in X$.
Then,
the \emph{TTS generated by $A_1\parallel A_2$} is
$\TTS{A_1\parallel A_2} = (S, s_0, \Sigma_1\cup\Sigma_2, {\rightarrow})$,
where
$S$ is the set of legal states,
$s_0=((\ell_1^0,\ell_2^0), v_{0})$, where $v_{0}$ maps each clock to 0,
and ${\rightarrow}$ is defined by
\begin{description}
  \item[Local action]
  $((\ell_1,\ell_2),v)\xrightarrow{a}((\ell'_1,\ell_2),v')$ iff
  $a\in\Sigma_1\setminus\Sigma_{2}^{\not\eps}$,
  $\ell_1\xrightarrow{g,a,r}\ell'_1$,
  $v \models g$, $v'= v[r]$ and $v' \models \inv_1(\ell'_1)$, and similarly
  for a local action in $\Sigma_2\setminus\Sigma_{1}^{\not\eps}$,
  \item[Synchronization]
  $((\ell_1,\ell_2),v)\xrightarrow{a}((\ell'_1,\ell'_2),v')$ iff $a\neq\eps$,
  $\ell_1\xrightarrow{g_1,a,r_1}\ell'_1$, \mbox{$\ell_2\xrightarrow{g_2,a,r_2}\ell'_2$},
  $v \models g_1\land g_2$, $v'= v[r_1\cup r_2]$ and
  $v' \models \inv_1(\ell'_1)\land\inv_2(\ell'_2)$,
  \item[Time delay] $\forall d\in \Reals,
  ((\ell_1,\ell_2),v)\xrightarrow{d}((\ell_1,\ell_2), v+d)$ iff
  $\forall d'\in[0,d], v+d' \models \inv_1(\ell_1)\land\inv_2(\ell_2)$.
\end{description}

A \emph{run} of an NTA  is an initial path in its TTS.
The semantics of a TA $A$ alone can also be given as a TTS denoted by
$\TTS{A}$ with only local actions and delay.
A TA is \emph{non-Zeno} iff for every infinite timed word $w$ generated by a
run, time diverges (\ie $\delta(w) = \infty$).
This is a common assumption for TA\@. In the sequel,
we always assume that the TA we deal with are non-Zeno.

\begin{rem}\label{rem:disjoint_clocksets}
  Let $A_1\parallel A_2$ be such that $X_1\cap X_2=\emptyset$. Then
  $\TTS{A_1}\x\TTS{A_2}$ is isomorphic to $\TTS{A_1}[A_2]$.
  This is not true in general when $X_1\cap X_2\neq\emptyset$.
  For example, in Fig.~\ref{fig:example1}, taking $b$ at time $0.5$ and
  $e$ at time 1 is possible in $\TTS{A_1}\x\TTS{A_2}$ but not in
  $\TTS{A_1}[A_2]$, since $b$ resets $x$ which is tested by
  $e$.
\end{rem}

\section{Need for Shared Clocks}
\label{sec:nsc_problem_setting}

\subsection{Problem Setting}
\begin{figure}[t]
  \centering
  \begin{tikzpicture}[node distance = 3cm,
                    initial where = left]
  \node[state, initial,initial text  = $A_1$] (q_0) [label=above:$x\leq2$] {};
  \node[state] (q_1) [right of=q_0] {};

  \node[state, initial,initial text  = $A_2$] (q_2) at (5,0) {};
  \node[state] (q_3) [right of=q_2] {};

  \path[->] (q_0) edge node {$x\geq1,a,\{x\}$} (q_1)
            (q_2) edge node {$x\leq2\land y\leq3,b$} (q_3);
  \end{tikzpicture}
  \caption{$A_2$ could avoid reading clock $x$ which belongs to $A_1$.}
  \label{fig:example0}
\end{figure}
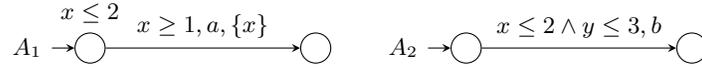
We are interested in detecting the cases where it is possible to avoid sharing
clocks, so that the model can be implemented using no other synchronization than
those explicitly described by common actions.

In this paper, we consider only the case of a network of two TA, $A_1\parallel
A_2$, such that $A_1$ does not read the clocks reset by $A_2$, and $A_2$
may read the clocks reset by $A_1$.
We want to know whether $A_2$ really needs to read these clocks,
or if another NTA $A'_1 \parallel A'_2$ could achieve the same
behavior as $A_1 \parallel A_2$ without using shared clocks.

A first remark is that our problem makes sense only if we insist on the
distributed nature of the system, made of two separate components. On the other
hand, if the composition operator is simply used as a convenient syntax for
describing a system that is actually implemented on a single sequential
component, then a simple product automaton would perfectly describe the system
and every clock becomes local.

So, let us consider the example of Fig.~\ref{fig:example0}, made of two TA,
supposed to describe two separate components. Remark that $A_2$ reads clock
$x$ which is reset by $A_1$. But a simple analysis shows that this reading
could be avoided: because of the condition on its clock $y$, $A_2$ can only
take transition $b$ before time $3$; but $x$ cannot reach value $2$
before time $3$, since it must be reset between time $1$ and $2$.
Thus, forgetting the condition on $x$ in $A_2$ would not change the
behavior of the system.

\subsection{Transmitting Information during Synchronizations}
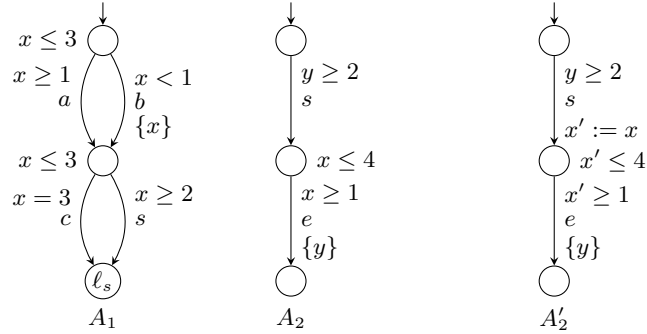
\begin{figure}[t]
  \centering
  \def\b{3.7}
  \begin{tikzpicture}[node distance = 1.6cm, initial where = above]
  \node[state, initial] (q_0) [label=left:$x\leq3$] {};
  \node[state] (q_1) [below of=q_0]
          [label=left:$x\leq3$] {};
  \node[none] (a_1) at (0,-\b) {$A_1$};
  \node[state] (q_4) [below of=q_1] {$\ell_s$};

  \node[state, initial] (q_2) at (2.5,0) {};
  \node[state] (q_3) [below of=q_2] [label=right:$x\leq4$] {};
  \node[state] (q_5) [below of=q_3] {};
  \node[none] (a_2) at (2.5,-\b) {$A_2$};

  \path[->] (q_0) edge [bend right] node [swap, pos=0.45] {$x\geq1$}
                                    node [swap] {$a$} (q_1)
            (q_1) edge [bend left] node [pos=0.45] {$x\geq2$}
                       node {$s$} (q_4)
                  edge [bend right,swap] node [pos=0.45] {$x=3$}
                       node {$c$} (q_4)
            (q_0) edge [bend left] node [pos=0.45]{$x<1$}
                                    node {$b$}
                                    node [pos=0.55] {$\{x\}$} (q_1)
            (q_2) edge node [pos=0.2]{$y\geq2$} node {$s$} (q_3)
            (q_3) edge node [pos=0.2] {$x\geq1$}
                                   node {$e$}
                                   node [pos=0.8] {$\{y\}$}(q_5);
  \begin{scope}[xshift=6cm]
    \node[state, initial] (q_2) {};
    \node[state] (q_3) [below of=q_2] [label=right:$x'\leq4$] {};
    \node[state] (q_5) [below of=q_3] {};
    \path[->] (q_2) edge node [pos=0.2]{$y\geq2$}
                         node {$s$}
                         node [pos=0.8] {$x':=x$}(q_3)
              (q_3) edge node [pos=0.2] {$x'\geq1$}
                         node {$e$}
                         node [pos=0.8] {$\{y\}$}(q_5);
  \node[none] (a_3) at (0,-\b) {$A'_2$};
  \end{scope}
\end{tikzpicture}
  \caption{$A_2$ reads $x$ which belongs to $A_1$ and $A'_2$ does not.}
  \label{fig:example1}
\end{figure}
Consider now the example of Fig.~\ref{fig:example1}. Here also $A_2$ reads clock
$x$ which is reset by $A_1$, and here also this reading could be avoided.
The idea is that $A_1$ could transmit the value of $x$ when
synchronizing, and afterwards any reading of $x$ in $A_2$ can be replaced
by the reading of a new clock $x'$ dedicated to storing the value of $x$
which is copied on the synchronization. Therefore $A_2$ can be replaced by
$A'_2$ pictured in Fig.~\ref{fig:example1}, while preserving the behavior of
the NTA, but also the behavior of $A_2$ w.r.t.\ $A_1$.

We claim that we cannot avoid reading $x$ without this copy of clock. Indeed,
after the synchronization, the maximal delay in the current location depends on
the exact value of $x$, and even if we find a mechanism to allow $A'_2$ to
move to different locations according to the value of $x$ at synchronization
time, infinitely many locations would be required (for example, if $s$ occurs
at time 2, $x$ may have any value in $(1,2]$).

\subsubsection*{Coding Transmission of Information}
In order to model the transmission of information during synchronizations, we
allow $A'_1$ and $A'_2$ to use a larger synchronization alphabet than
$A_1$ and $A_2$. This allows $A'_1$ to transmit discrete information,
like its current location, to $A'_2$.

But we saw that $A'_1$ also needs to transmit the exact value of its clocks,
which requires a more general mechanism than the simple clock resets.
For this we allow an automaton to copy its neighbor's clocks into local clocks
during synchronizations. This is denoted as updates of the form $x':=x$ in
$A'_2$ (see Fig.~\ref{fig:example1}).
This feature is a bit unusual but has already been studied: it is a restricted
class of updatable timed automata as defined in \cite{upd}.
Moreover, as shown in \cite{upd}, the class we consider, without comparisons of
clocks and with only equalities in the updates
is not more expressive than classical TA for the sequential semantics
(any updatable TA of the class is bisimilar to a classical TA), and the
emptiness problem is \textsf{PSPACE}-complete, as in the case of classical TAs.

\subsubsection*{Semantics}
$\TTS{A_1}[A_2]$ can be defined as previously,
with the difference that the synchronizations are now defined by:
$((\ell_1,\ell_2),v)\xrightarrow{a}((\ell_1',\ell_2'),v')$
iff $\ell_1\xrightarrow{g_1,a,r_1}_1\ell_1'$,
$\ell_2\xrightarrow{g_2,a,r_2,u}_2\ell_2'$ where $u$ is a partial function
from $X_2$ to $X_1$, \mbox{$v\models g_1\land g_2$},
\mbox{$v'=(v[r_1\cup r_2])[u]$},
and $v'\models\inv(\ell_1')\land\inv(\ell_2')$. The valuation $v[u]$ is defined by
$v[u](x)=v(u(x))$ if $u(x)$ is defined, and $v[u](x)=v(x)$ otherwise.

Here, we choose to apply the reset $r_1\cup r_2$ before the update $u$,
because we are interested in sharing the state reached in $A_1$ after the
synchronization, and $r_1$ may reset some clocks in $C_1\subseteq X_1$.

\subsection{Towards a Formalization of the Problem}\label{subsec:towards}

We want to know whether $A_2$ really needs to read the clocks reset by
$A_1$, or if another NTA $A'_1 \parallel A'_2$ could achieve the same
behavior as $A_1 \parallel A_2$ without using shared clocks. It remains to
formalize what we mean by ``having the same behavior'' in this context.

\begin{figure}[t]
  \centering
  \def\a{0.75}
\def\b{1.5}
\def\c{4}
\begin{tikzpicture}[node distance = 2cm, initial where = above]
  \tikzstyle{every label}=[label distance=-2pt]
  \node[state, initial] (q_0) [label=left:$x\leq1$] {$p_{0}$};
  \node[state] (q_1) at (-\a,-\b) {$p_{1}$};
  \node[state] (q_2) at (\a,-\b) {$p_{2}$};
  \node[none] (a_1) at (0,-2.3*\b) {$A_1$};

  \begin{scope}[xshift=\c cm]
  \node[state, initial] (q_02) [label=left:$y\leq2$] {$q_{0}$};
  \node[state] (q_12) at (-2*\a,-\b) [label=right:$y\leq2$] {$q_{1}$};
  \node[state] (q_22) at (2*\a,-\b) [label=left:$y\leq2$] {$q_{2}$};
  \begin{scope}[xshift=-2*\a cm]
  \node[state] (q_32) at (-0.45*\a,-2*\b) {$q_{3}$};
  \node[state] (q_42) at (0.45*\a,-2*\b) {$q_{4}$};
  \end{scope}
  \begin{scope}[xshift=2*\a cm]
  \node[state] (q_52) at (-0.45*\a,-2*\b) {$q_{5}$};
  \node[state] (q_62) at (0.45*\a,-2*\b) {$q_{6}$};
  \end{scope}
  \node[none] (a_2) at (0,-2.3*\b) {$A_2$};
  \end{scope}

  \begin{scope}[xshift=2.1*\c cm]
  \node[state, initial] (q_03) [label=left:$y\leq2$] {};
  \node[state] (q_13) at (-1.6*\a,-\b) [label=right:$y\leq2$] {$r_{1}$};
  \node[state] (q_23) at (1.6*\a,-\b) [label=left:$y\leq2$] {$r_{2}$};
  \node[state] (q_33) at (-1.6*\a,-2*\b) {};
  \node[state] (q_43) at (1.6*\a,-2*\b) {};
  \node[none] (a_3) at (0,-2.3*\b) {$A'_2$};
  \end{scope}

  \tikzstyle{every node}=[inner sep=1.5pt]
  \path[->] (q_0) edge [swap] node [pos=0.25] {$x=1$}
                       node {$d$} (q_1)
                  edge node [pos=0.25] {$x=1$}
                       node {$e$}
                       node [pos=0.75] {$\{x\}$} (q_2)
           (q_02) edge [swap] node [pos=0.45] {$y=2$}
                       node [pos=0.65] {$c$} (q_12)
                  edge node [pos=0.45] {$y=2$}
                       node [pos=0.65] {$c$} (q_22)
           (q_12) edge [swap] node [pos=0.45] {$y=2\mathop{\land}$}
                       node [pos=0.65] {$x=1$}
                       node [pos=0.9] {$a$} (q_32)
                  edge node [pos=0.45] {$y=2\mathop{\land}$}
                       node [pos=0.65] {$x=2$}
                       node [pos=0.9] {$b$} (q_42)
           (q_22) edge [swap] node [pos=0.45] {$y=2\mathop{\land}$}
                       node [pos=0.65] {$x=1$}
                       node [pos=0.9] {$b$} (q_52)
                  edge node [pos=0.45] {$y=2\mathop{\land}$}
                       node [pos=0.65] {$x=2$}
                       node [pos=0.9] {$a$} (q_62)
           (q_03) edge [swap] node [pos=0.45] {$y=2$}
                       node [pos=0.65] {$c$} (q_13)
                  edge node [pos=0.45] {$y=2$}
                       node [pos=0.65] {$c$}  (q_23)
           (q_13) edge node [pos=0.25] {$y=2$}
                       node {$a$} (q_33)
           (q_23) edge [swap] node [pos=0.25] {$y=2$}
                       node {$b$} (q_43);
\end{tikzpicture}
  \caption{$A_2$ needs to read the clocks of $A_1$ and
  $\TTS{A_1}[A_2]\wtb\TTS{A_1}[A'_2]$. 
  \label{fig:bisim}}
\end{figure}
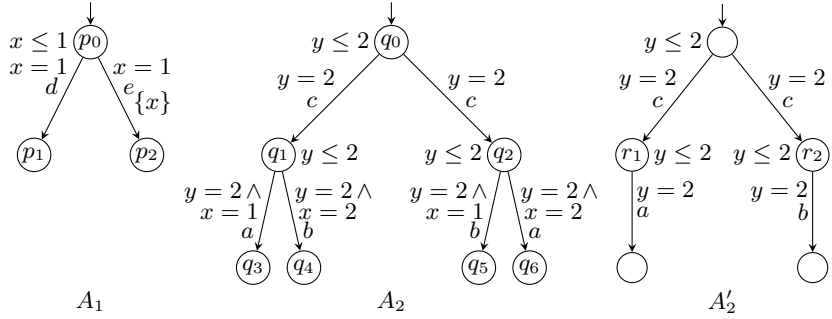

First, we impose that the locality of actions is preserved, \ie $A'_1$ uses
the same set of local actions as $A_1$, and similarly for $A'_2$ and
$A_2$. For the synchronizations, we have explained earlier why we allow
$A'_1$ and $A'_2$ to use a larger synchronization alphabet than $A_1$ and
$A_2$. The correspondence between both alphabets will be done by a mapping
$\psi$ (this point will be refined later).

Now we have to ensure that the behavior is preserved. The first idea that comes
to mind is to impose bisimulation between $\psi(\TTS{A'_1}[A'_2])$
(\ie $\TTS{A'_1}[A'_2]$ with synchronization actions relabeled by
$\psi$) and $\TTS{A_1}[A_2]$. But this is not sufficient, as illustrated by
the example of Fig.~\ref{fig:bisim} (where $\psi$ is the identity).
Intuitively $A_2$ needs to read $x$ when in $q_{1}$ (and similarly in
$q_{2}$) at time 2, because this reading determines whether it will
perform $a$ or $b$, and the value of $x$ cannot be inferred from its local
state given by $q_{1}$ and the value of $y$. Anyway $\TTS{A_1}[A'_2]$
is bisimilar to $\TTS{A_1}[A_2]$, and $A'_2$ does not read $x$.
For the bisimulation relation $\R$, it is sufficient to impose
$(p_{1},q_{1})\R(p_{1},r_{1})$ and
$(p_{2},q_{1})\R(p_{2},r_{2})$.

What we see here is that, from the point of view of $A_2$ and $A'_2$,
these two automata do not behave the same. As a matter of fact, when  $A_2$
fires one edge labeled by $c$, it has not read $x$ yet, and there is still a
possibility to fire $a$ or $b$, whereas when $A'_2$ fires one edge labeled
by $c$, there is no more choice afterwards. Therefore we need a relation
between $A'_2$ and $A_2$, and in the general case,
a relation between $A'_1$ and $A_1$ also.

\section{Contextual Timed Transition Systems}\label{sec:TTS}

As we are interested in representing a partial view of one of the components, we
need to introduce another notion, that we call \emph{contextual timed transition
system}. This resembles the powerset construction used in game theory to
capture the knowledge of an agent about another agent~\cite{Reif}.

\subsubsection*{Notations}
$\sync=\Sigma_{1}^{\not\eps}\cap\Sigma_{2}^{\not\eps}$ denotes the set of common actions.
$Q_1$ denotes the set of states of $\TTS{A_1}$.
When $s=((\ell_1,\ell_2),v)$ is a state of $\TTS{A_1}[A_2]$, we also write
\mbox{$s=(s_1,s_2)$}, where $s_1=(\ell_1,v_{|X_1})$ is in $Q_1$, and
$s_2=(\ell_2,v_{|X_2\setminus X_1})$, where $v_{|X}$ is $v$ restricted to
$X$.

\begin{defi}[$\UR(s)$]
Let \mbox{$\TTS{A_1}=(Q_1,s_0,\Sigma_1,{\rightarrow}_1)$} and $s\in Q_1$.
The set of states of $A_1$ reachable from $s$ by local actions in 0 delay
(and therefore not observ\-able by $A_2$) is denoted by $\UR(s)$ (for
Unobservably Reachable) and defined as
\[UR(s) = \{s'\in Q_1\mid
\exists w\in\TW{\Sigma_1\setminus\Sigma_{2}^{\not\eps}}: s\xrightarrow{w}_1s'\}\;.\]
\end{defi}

\subsection{Contextual TTS}

\subsubsection*{Contextual States}
The states of this contextual TTS are called \emph{contextual states}. They can
be regarded as possibly infinite sets of states of  $\TTS{A_1}[A_2]$ for which
$A_2$ is in the same location and has the same valuation over $X_2\setminus X_1$.
$A_2$ may not be able to distinguish between some states
$(s_1,s_2)$ and $(s'_1,s_2)$. In $\TTS[A_1]{A_2}$, these states are
grouped into the same contextual state.
However, when $X_2\cap X_1\neq\emptyset$, it may happen that $A_2$ is able
to perform a local action or delay from $(s_1,s_2)$ and not from
$(s'_1,s_2)$, even if these states are grouped in a same contextual state.

\begin{defi}[Contextual TTS]
Let \mbox{$\TTS{A_1}[A_2]=(Q, q_0, \Sigma_1\cup\Sigma_2, {\Rightarrow})$}.
Then, the \emph{TTS of $A_2$ in the context of $A_1$}, denoted by
$\TTS[A_1]{A_2}$, is the TTS
$(S,s_0,(\Sigma_2\setminus\sync)\cup(\sync\times Q_1),{\rightarrow})$, where
\begin{itemize}
  \item $S=\{(S_1,s_2)\mid \forall s_1\in S_1,
  (s_1,s_2)\in Q\}$, 
  \item $s_0=(S_1^0,s_2^0)$, s.t.\ $(s_1^0,s_2^0)=q_0$ and
  $S_1^0=\UR(s_1^0)$,
  \item ${\rightarrow}$ is defined by
  \begin{itemize}
    \item Local action: for any $a\in\Sigma_2\setminus\sync$,
    $(S_1,s_2)\xrightarrow{a}(S'_1,s'_2)$
    iff $\exists s_1\in S_1: (s_1,s_2)\xRightarrow{a}(s_1,s'_2)$,
    and $S'_1=\{s_1\in S_1\mid(s_1,s_2)\xRightarrow{a}(s_1,s'_2)\}$
    \item Synchronization: for any $(a,s'_1)\in\sync\times Q_1$,
    $(S_1,s_2)\xrightarrow{a,s'_1}(\UR(s'_1),s'_2)$
    iff $\exists s_1\in S_1: (s_1,s_2)\xRightarrow{a}(s'_1,s'_2)$
    \item Local delay: for any $d\in\Reals$,
    $(S_1,s_2)\xrightarrow{d}(S'_1,s'_2)$
    iff $\exists s_1\in S_1, \rho\in \paths{\Sigma_1\setminus\Sigma_{2}^{\not\eps},d}:
    (s_1,s_2)\xRightarrow{\rho}(s'_1,s'_2)$, and  $S'_1=\{s'_1\mid
    \exists s_1\in S_1, \rho\in \paths{\Sigma_1\setminus\Sigma_{2}^{\not\eps},d}:
    (s_1,s_2)\xRightarrow{\rho}(s'_1,s'_2)\}$
  \end{itemize}
\end{itemize}
\end{defi}

\noindent For example, consider $A_1$ and $A_2$ of Fig.~\ref{fig:bisim}. The initial
state is $\big(\{(p_{0},0)\},(q_{0},0)\big)$. From this contextual
state, it is possible to delay $2$ time units and reach the contextual
state $\big(\{(p_{1},2),(p_{2},1)\},(q_{0},2)\big)$. Indeed,
during this delay, $A_1$ has to perform either $e$ and reset $x$, or
$d$. Now, from this contextual state, we can take an edge labeled by $c$,
and reach $\big(\{(p_{1},2),(p_{2},1)\},(q_{1},2)\big)$. Lastly,
from this new state, $a$ can be fired, because it is enabled by
$((p_{2},1),(q_{1},2))$ in the TTS of the NTA, and the reached
contextual state is $\big(\{(p_{2},1)\},(q_{3},2)\big)$.

\subsubsection*{Unrestricted Contextual TTS}
We say that there is no restriction in $\TTS[A_1]{A_2}$ if whenever a local
step is possible from a reachable contextual state, then it is possible from all
the states $(s_1,s_2)$ that are grouped into this contextual state.
In the example above, there is a restriction in $\TTS[A_1]{A_2}$ because
we have seen that $a$ is enabled only by $((p_{2},1),(q_{1},2))$,
and not by all states merged in
$\big(\{(p_{1},2),(p_{2},1)\},(q_{1},2)\big)$.
Formally, we use the predicate $\noRest[A_1]{A_2}$ defined as follows.

\begin{defi}[\text{$\noRest[A_1]{A_2}$}]
The predicate $\noRest[A_1]{A_2}$ holds iff for any reachable state
$(S_1,s_2)$ of $\TTS[A_1]{A_2}$, both
\begin{itemize}
  \item $\forall a\in\Sigma_2\setminus\sync,
    (S_1,s_2)\xrightarrow{a}(S'_1,s'_2) \iff
    \forall s_1\in S_1, (s_1,s_2)\xRightarrow{a}(s_1,s'_2)$, and
  \item $\forall d\in\Reals,(S_1,s_2)\xrightarrow{d} \iff
    \forall s_1\in S_1, \exists \rho\in \paths{\Sigma_1\setminus\Sigma_{2}^{\not\eps},d}:
    (s_1,s_2)\xRightarrow{\rho}$
\end{itemize}
\end{defi}
\begin{rem}\label{rem:noread_norest}
  If $A_2$ does not read $X_1$, then there is no restriction in
  $\TTS[A_1]{A_2}$.
\end{rem}

\subsubsection*{Sharing of Information During Synchronizations}
Later we assume that during a synchronization, $A_1$ is allowed to transmit all
its state to $A_2$, that is why, in $\TTS[A_1]{A_2}$, we distinguish the
states reached after a synchronization according to the state reached in
$A_1$. We also label the synchronization edges by a pair $(a,s_1)\in
\sync\times Q_1$ where $a$ is the action and $s_1$ the state reached in
$A_1$. 

For the sequel, let $\TTS[Q_1]{A_1}$ (resp. $\TTS[Q_1]{A_1}[A_2]$)
denote $\TTS{A_1}$ (resp. $\TTS{A_1}[A_2]$) where the synchronization edges
are labeled by $(a,s_1)$, where $a\in\sync$ is the action, and $s_1$ is
the state reached in $A_1$.

We can now state a nice property of unrestricted contextual TTS that is similar
to the distributivity of TTS over the composition when considering TA with
disjoint sets of clocks (see Remark~\ref{rem:disjoint_clocksets}). We say that a TA
is \emph{deterministic} if it has no $\eps$-transition and for any location
$\ell$ and action $a$, there is at most one edge labeled by $a$ from
$\ell$.

\begin{lem}
  \label{lem:distrib}
  If there is no restriction in $\TTS[A_1]{A_2}$, then
  $\TTS[Q_1]{A_1}\x\TTS[A_1]{A_2}\stb\TTS[Q_1]{A_1}[A_2]$.
  Moreover, when $A_2$ is deterministic, this condition becomes necessary.
\end{lem}
\begin{figure}[t]
  \centering
  \def\a{1.75}
\def\b{0.4}
\def\c{4}
\begin{tikzpicture}[node distance = 2cm]
  \tikzstyle{every label}=[label distance=-2pt]

  \node[state, initial, initial text=$A_1$] (p_0) {};

  \begin{scope}[xshift=\c cm]
  \node[state, initial, initial text=$A_2$] (q_0) {};
  \node[state] (q_1) at (\a,\b) {};
  \node[state] (q_2) at (\a,-\b) {};
  \end{scope}

  \tikzstyle{every node}=[inner sep=1.5pt]
  \path[->] (q_0) edge node [sloped,pos=0.9] {$x<1,a$} (q_1)
                  edge node [swap,sloped,pos=0.5] {$a$} (q_2)
            (p_0) edge [loop right] node {$x\geq1,b,\{x\}$} (p_0);
\end{tikzpicture}
  \caption[$A_2$ is non-deterministic and the reciprocal of
  Lemma~\ref{lem:distrib} does not hold.]
  {
  $\TTS[Q_1]{A_1}\x\TTS[A_1]{A_2}\stb\TTS[Q_1]{A_1}[A_2]$,
  although there is a restriction in $\TTS[A_1]{A_2}$\label{fig:counterex}}
\end{figure}
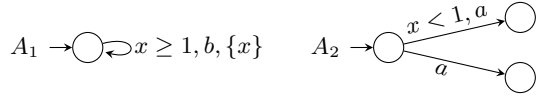
The example of Fig.~\ref{fig:counterex} shows that the reciprocal does not hold
when $A_2$ is not deterministic.
In order to prove Lemma~\ref{lem:distrib}, we first present two propositions.
The first one relates the reachable states of $\TTS[A_1]{A_2}$ with those of
$\TTS[Q_1]{A_1}\x\TTS[A_1]{A_2}$.

\begin{prop}\label{prop:reachable}\hfill
\begin{enumerate}
  \item For any reachable state $(S_1,s_2)$ of $\TTS[A_1]{A_2}$,\\
  $s_1\in S_1\implies(s_1,(S_1,s_2))\text{ is a reachable state of }
  \TTS[Q_1]{A_1}\x\TTS[A_1]{A_2}$
  \item $\noRest[A_1]{A_2}$ iff\\
  $\begin{array}{@{\quad}l}
    \text{for any reachable state }(S_1,s_2)\text{ of }\TTS[A_1]{A_2},\\
    s_1\in S_1\iff(s_1,(S_1,s_2))\text{ is a reachable state of }
    \TTS[Q_1]{A_1}\x\TTS[A_1]{A_2}
  \end{array}$
  \end{enumerate}
\end{prop}
\proof
  (1) For any reachable state $(S_1,s_2)$, let us denote by $P(S_1,s_2)$ the
  fact that for any $s_1\in S_1$, $(s_1,(S_1,s_2))$ is reachable in
  $\TTS[Q_1]{A_1}\x\TTS[A_1]{A_2}$. We give a recursive proof. First, the
  initial state $(S_1^0,s_2^0)$ satisfies $P(S_1^0,s_2^0)$ because for any
  $s_1\in S_1^0=\UR(s_1^0)$, $\exists w\in\TW{\Sigma_1\setminus\Sigma_{2}^{\not\eps}}:
  s_1^0\xrightarrow{w}_1 s_1$ and hence $(s_1^0,(S_1^0,s_2^0))\xrightarrow{w}
  (s_1,(S_1^0,s_2^0))$.
  Then, assume some reachable state $(S_1,s_2)$ of $\TTS[A_1]{A_2}$
  satisfies $P(S_1,s_2)$ and show that any state $(S'_1,s'_2)$ reachable in
  one step from $(S_1,s_2)$ also satisfies $P(S'_1,s'_2)$.
  There can be three kinds of steps from $(S_1,s_2)$ in $\TTS[A_1]{A_2}$.
  \begin{enumerate}
  \item If for some $a\in \Sigma_2\setminus\sync$,
    $(S_1,s_2)\xrightarrow{a}(S'_1,s'_2)$, then for any
    \mbox{$s'_1\in S'_1\subseteq S_1$},
    $(s'_1,(S_1,s_2))\xrightarrow{a}(s'_1,(S'_1,s'_2))$,
    \ie $P(S'_1,s'_2)$ holds.
  \item If for some $(a,s'_1)\in \sync\times Q_1$,
    $(S_1,s_2)\xrightarrow{a,s'_1}(S'_1,s'_2)$, then $S'_1=\UR(s'_1)$, and
    for some \mbox{$s_1\in S_1$},
    $(s_1,(S_1,s_2))\xrightarrow{a,s'_1}(s'_1,(S'_1,s'_2))$.
    By the same reasoning as for $(S_1^0,s_2^0)$, for any
    $s_1''\in S'_1=\UR(s'_1)$, $\exists w\in\TW{\Sigma_1\setminus\Sigma_{2}^{\not\eps}}:
    (s'_1,(S'_1,s'_2))\xrightarrow{w}(s''_1,(S'_1,s'_2))$.
    Hence $P(S'_1,s'_2)$ holds.
  \item If for some $d\in \Reals$,
    $(S_1,s_2)\xrightarrow{d}(S'_1,s'_2)$, then
    $\exists d_1\leq d: (S_1,s_2)\xrightarrow{d_1}(S_1^1,s_2^1)\land
    \exists s_1^1\in S_1^1,s_1\in S_1:
    (s_1,s_2)\xRightarrow{d_1}(s_1^1,s_2^1)$,
    that is $(s_1^1,(S_1^1,s_2^1))$ is reachable, and by time-determinism,
    $(S_1^1,s_2^1)\xrightarrow{d-d_1}(S'_1,s'_2)$.
  \end{enumerate}
  For the third case, take $d_1$ small enough (but strictly positive) so that
  $S_1^1=\{s_1'\mid\exists s_1\in S_1:
  (s_1,s_2)\xRightarrow{d_1}(s_1^1,s_2^1) \land s'_1\in\UR(s_1^1)\}$.
  That is, after some local actions that take no time, $A_1$ is able to
  perform a delay $d_1$ during which no local action is enabled (such $d_1$
  exists because  of the non-zenoness assumption). With such $d_1$, any state
  $s_1'\in S_1^1$ is such that $s'_1\in\UR(s_1^1)$ for some $s_1^1$ so
  that $(s_1^1,(S_1^1,s_2^1))$ is reachable. Therefore,
  $\exists w\in\TW{\Sigma_1\setminus\Sigma_{2}^{\not\eps}}:(s_1^1,(S_1^1,s_2^1))
  \xrightarrow{w}(s'_1,(S_1^1,s_2^1))$ and hence $P(S_1^1,s_2^1)$ holds.

  Since $A_1$ is not Zeno, any delay in $\TTS[A_1]{A_2}$ can be cut into a
  finite number of such smaller global delays. Hence, for any $(S_1,s_2)$ that
  satisfies $P(S_1,s_2)$, for any $d\in \Reals$ such that
  $(S_1,s_2)\xrightarrow{d}(S'_1,s'_2)$, $P(S'_1,s'_2)$ holds.

  (2, $\Rightarrow$) (1) already gives that
  $\forall s_1\in S_1$, $(s_1,(S_1,s_2))$ is a reachable state.
  So it remains to prove that, when $\noRest[A_1]{A_2}$,
  if $(s_1,(S_1,s_2))$ is a reachable state, then
  $s_1\in S_1$. We say that a reachable state $s=(s_1,(S_1,s_2))$ satisfies
  $P(s)$ iff $s_1\in S_1$.

  Assume $\noRest[A_1]{A_2}$ and $s=(s_1,(S_1,s_2))$ is a reachable
  state that satisfies $P(s)$.
  Then, any state $s'$ reachable in one step from $s$ by some local action
  or delay $a\in(\Sigma_1\cup\Sigma_2)\setminus\sync\cup\Reals$ or by some synchronization
  $(a,s'_1)\in\sync\times Q_1$ matches one of the following cases:
  \begin{itemize}
    \item if $a\in\Sigma_1\setminus\Sigma_{2}^{\not\eps}$, then $s'=(s'_1,(S_1,s_2))$
    such that  $s'_1\in\UR(s_1)\subseteq S_1$ (by construction,
    $s_1\in S_1\implies\UR(s_1)\subseteq S_1$),
    \item if $a\in\Sigma_2\setminus\Sigma_1$, then $s'=(s_1,(S_1,s'_2))$,
    \item if $a\in\Reals$, then $s'=(s'_1,(S'_1,s'_2))$,
    where $s'_1$ such that $(s_1,s_2)\xRightarrow{a}(s'_1,s'_2)$ is in
    $S'_1=\{q'_1\mid\exists q_1\in S_1,\rho\in\paths{\Sigma_1\setminus\Sigma_{2}^{\not\eps},a}:
    (q_1,s_2)\xRightarrow{\rho}(q'_1,s'_2)\}$,
    \item if $(a,s'_1)\in(\sync\times Q_1)$, then
    $s'=(s'_1,(\UR(s'_1),s'_2))$.
  \end{itemize}
  Therefore, any state $s'$ reached in one step from $s$ also satisfies
  $P(s')$, and recursively, since the initial state
  $s_0=(s_1^0,(\UR(s_1^0),s_2^0))$ satisfies $P(s_0)$, any reachable state
  $s$ of $\TTS[Q_1]{A_1}\x\TTS[A_1]{A_2}$ satisfies $P(s)$.

  (2, $\Leftarrow$) 
  By contradiction, assume there is a restriction in state $(S_1,s_2)$ for
  local delay or action $a\in(\Sigma_2\setminus\Sigma_1)\cup\Reals$ \ie
  $a$ is possible from some state $(s'_1, s_2)$ but not from another state
  $(s_1,s_2)$ such that $s'_1,s_1\in S_1$. Then, after performing $a$ from
  $(s_1,(S_1,s_2))$, that is reachable according to
  Proposition~\ref{prop:reachable}, we reach state $(s_1,(S'_1,s'_2))$ such
  that $s_1\notin S'_1$.\qed

\begin{prop}\label{prop:uniqueness}
  If $\noRest[A_1]{A_2}$ then,
  for any timed word $w$ over $(\Sigma_2\setminus\sync)\cup
  (\sync\times Q_1)$, there exists at most one $S_1$ such that, for some $s_2$,
  $(S_1^0,s_2^0)\xrightarrow{w}(S_1,s_2)$ in $\TTS[A_1]{A_2}$
  (\ie $S_1$ is uniquely determined by $w$, whatever the structure of
  $A_2$).
\end{prop}
\proof
Assuming $\noRest[A_1]{A_2}$, we show that, for any $(S_1^1,s_2^1)$ reachable
in $\TTS[A_1]{A_2}$, for any action or delay in
$(\Sigma_2\setminus\sync)\cup(\sync\times Q_1)\cup\Reals$,
there is at most one $S_1$ such that, for some $s_2$, $(S_1,s_2)$ is a
successor of $(S_1^1,s_2^1)$ by this action.

Indeed, by construction, and since there is no restriction,
\begin{itemize}
  \item any successor of $(S_1^1,s_2^1)$ by a local action is of the form
  $(S_1^1,s_2')$,
  \item any successor of $(S_1^1,s_2^1)$ by a synchronization $(a,s'_1)$ is
  of the form $(\UR(s'_1),s_2')$,
  \item any successor of $(S_1^1,s_2^1)$ by a delay $d$ is
  of the form $(S_1,s_2')$ with $S_1=\{s'_1\mid
  \exists \rho\in\paths{\Sigma_1\setminus\Sigma_{2}^{\not\eps},d},s_1\in S_1^1:
  s_1\xrightarrow{\rho}_1 s'_1\}$.
\end{itemize}
Therefore, for any possible action or delay, $S_1$ does not depend on the
state of $A_2$, and is uniquely determined by this action or delay.

Since $(S_1^0,s_2^0)$ is unique, for any timed word $w$ over
$(\Sigma_2\setminus\sync)\cup(\sync\times Q_1)$, either $w$ does not
describe a valid path in $\TTS[A_1]{A_2}$, or there exists a unique $S_1$
such that for some $s_2$,
$(S_1^0,s_2^0)\xrightarrow{w}(S_1,s_2)$ in $\TTS[A_1]{A_2}$.\qed

We can now prove Lemma~\ref{lem:distrib}.
\proof[Proof of Lemma~\ref{lem:distrib}]
  Assume $\noRest[A_1]{A_2}$, and define relation
  $\R$ as $(s_1,(S_1,s_2))\R (s'_1,s'_2)\iffdef s_1=s'_1 \land s_2=s'_2$,
  for any reachable states
  $(s_1,(S_1,s_2))$ of $\TTS[Q_1]{A_1}\x\TTS[A_1]{A_2}$ and
  $(s'_1,s'_2)$ of $\TTS[Q_1]{A_1}[A_2]$.
  By Proposition~\ref{prop:reachable}, since $(s_1,(S_1,s_2))$
  is reachable, $s_1\in S_1$.
  We show that $\R$ is a strong timed bisimulation.

  First, the initial states
  are $\R$-related: $(s_1^0,(S_1^0,s_2^0))\R(s_1^0,s_2^0)$.
  Then, if $(s_1,(S_1,s_2))\R (s'_1,s'_2)$, four kinds of steps are possible:
  \begin{itemize}
    \item if for some $a\in \Sigma_1\setminus\Sigma_{2}^{\not\eps}$,
    $(s_1,(S_1,s_2))\xrightarrow{a}(s'_1,(S_1,s_2))$, then
    $(s_1,s_2)\xRightarrow{a}(s'_1,s_2)$ and $(s'_1,(S_1,s_2))\R(s'_1,s_2)$,
    and conversely.
    \item if for some $a\in \Sigma_2\setminus\Sigma_1$,
    $(s_1,(S_1,s_2))\xrightarrow{a}(s_1,(S_1,s'_2))$, then, $\forall s_{11}\in S_1$,
    $(s_{11},s_2)\xRightarrow{a}(s_{11},s'_2)$ (because $\noRest[A_1]{A_2}$),
    and in particular, $(s_1,s_2)\xRightarrow{a}(s_1,s'_2)$ and
    $(s_1,(S_1,s'_2))\R(s_1,s'_2)$, and conversely.
    \item if for some $(a,s'_1)\in \sync\times Q_1$,
    $(s_1,(S_1,s_2))\xrightarrow{a,s'_1}(s'_1,(S'_1,s'_2))$, then
    $(s_1,s_2)\xRightarrow{a,s'_1}(s'_1,s'_2)$ and $(s'_1,(S'_1,s'_2))\R(s'_1,s'_2)$,
    and conversely.
    \item if for some $d\in \Reals$,
    $(s_1,(S_1,s_2))\xrightarrow{d}(s'_1,(S'_1,s'_2))$, then
    $(s_1,s_2)\xRightarrow{d}(s'_1,s'_2)$ (because $\noRest[A_1]{A_2}$),
    and $(s'_1,(S'_1,s'_2))\R(s'_1,s'_2)$, and conversely.
  \end{itemize}

  \noindent Now assume $A_2$ is deterministic.
  Let relation $\R$ 
  be a strong timed bisimulation between $\TTS[Q_1]{A_1}\x\TTS[A_1]{A_2}$ and
  $\TTS[Q_1]{A_1}[A_2]$. 

  By contradiction, assume there is a restriction in $\TTS[A_1]{A_2}$. Then
  there is a reachable state $(S_1,s_2)$ of $\TTS[A_1]{A_2}$, and a local
  delay or action $a\in (\Sigma_2\setminus\Sigma_1)\cup\Reals$ such that, for
  some $s_1,s'_1\in S_1$, $(s_1,s_2)$ enables $a$ in
  $\TTS[Q_1]{A_1}[A_2]$, whereas $(s'_1,s_2)$ does not.

  By definition of a bisimulation, there also exist two states
  $(p_1,(P_1,p_2))$ and $(p'_1,(P'_1,p'_2)$ such that
  $(p_1,(P_1,p_2))\R(s_1,s_2)$ and $(p'_1,(P'_1,p'_2))\R(s'_1,s_2)$. That
  is, in particular, $(p'_1,(P'_1,p'_2))$ does not enable $a$.
  Moreover, these states can be chosen so that they are reached by the same
  timed word over $(\Sigma_2\setminus\sync)\cup(\sync\times Q_1)$, and since
  $A_2$ is deterministic, $p_2=p'_2=s_2$.

  Now, we can assume that $(S_1,s_2)$ is chosen so that it is the first state
  with a restriction along an initial path. Then, the paths to $(P_1,s_2)$ and
  $(P'_1,s_2)$ generate the same timed word over
  $(\Sigma_2\setminus\sync)\cup(\sync\times Q_1)$, and by
  Proposition~\ref{prop:uniqueness}, $P_1=P'_1=S_1$. 

  Therefore, we have shown the existence of a state $(p'_1,(S_1,s_2))$ in
  $\TTS[Q_1]{A_1}\x\TTS[A_1]{A_2}$ that does not enable $a$, which means
  that $(S_1,s_2)$ does not enable $a$ in $\TTS[A_1]{A_2}$.
  This contradicts the fact that there exists $s_1\in S_1$ such that
  $(s_1,s_2)$ enables $a$.\qed

We are now in condition to formalize our problem.

\subsection{Need for Shared Clocks Revisited}\label{subsec:nsc}                %
We have argued in Section~\ref{subsec:towards} that the existence of a
NTA $A'_1 \parallel A'_2$ without shared clocks and such that
$\psi(\TTS[Q'_1]{A'_1}[A'_2])\wtb\TTS[Q_1]{A_1}[A_2]$ is not sufficient to
capture the idea that $A_2$ does not need to read the clocks of $A_1$.
We are now equipped to define the relations we want to impose
on the separate components, namely
$\psi(\TTS[Q'_1]{A'_1})\wtb\TTS[Q_1]{A_1}$ and
$\psi(\TTS[A'_1]{A'_2})\wtb\TTS[A_1]{A_2}$.
And since we have seen the importance of labeling the synchronization
actions in contextual TTS by labels in $\sync\times Q_1$ rather than in
$\sync$, the correspondence between the synchronization labels of $A'_1
\parallel A'_2$ with those of $A_1 \parallel A_2$ is now done by a mapping
$\psi: \sync'\times Q'_1 \rightarrow \sync\times Q_1$.

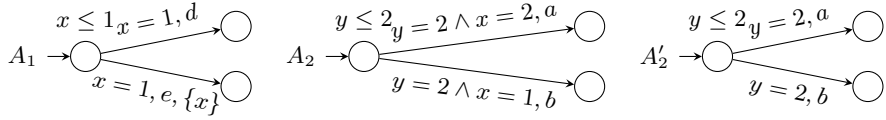
\begin{figure}[t]
  \centering
  \def\b{-0.4}
\def\c{2}
\def\d{1}
\def\a{\c+\d}
\begin{tikzpicture}[node distance = 2cm]
  \node[state, initial, initial text  = $A_1$] (q_0) [label=above:$x\leq1$] {};
  \node[state] (q_1) at (\c,\b) {};
  \node[state] (q_2) at (\c,-\b) {};

  \begin{scope}[xshift=4.2cm-\d cm/2]
  \node[state, initial, initial text  = $A_2$] (q_02) [label=above:$y\leq2$] {};
  \node[state] (q_12) at (\a,\b) {};
  \node[state] (q_22) at (\a,-\b) {};
  \end{scope}

  \begin{scope}[xshift=8.4cm]
  \node[state, initial, initial text  = $A'_2$] (q_03) [label=above:$y\leq2$] {};
  \node[state] (q_13) at (\c,\b) {};
  \node[state] (q_23) at (\c,-\b) {};
  \end{scope}

  \path[->] (q_0) edge node [sloped,above] {$x=1,d$} (q_2)
                  edge node [sloped,below] {$x=1,e,\{x\}$} (q_1)
           (q_02) edge node [sloped,above] {$y=2\land x=2, a$} (q_22)
                  edge node [sloped,below] {$y=2\land x=1, b$} (q_12)
           (q_03) edge node [sloped,above] {$y=2, a$} (q_23)
                  edge node [sloped,below] {$y=2, b$} (q_13);
\end{tikzpicture}
  \caption{$A_2$ needs to read the clocks of $A_1$ and
  $\TTS[A_1]{A_2}\wtb\TTS[A_1]{A'_2}$. 
  \label{fig:bisim2}}
\end{figure}

This settles the problem of the example of Fig.~\ref{fig:bisim} where
$\TTS[A_1]{A'_2}\not\wtb\TTS[A_1]{A_2}$ (here $A'_1=A_1$), but as shown in
Fig.~\ref{fig:bisim2}, a problem remains.
In this example, we can see that $A_2$
needs to read clock $x$ of $A_1$ to know whether it has to perform $a$
or $b$ at time 2, and yet $\TTS[A_1]{A_2}\wtb\TTS[A_1]{A_2'}$
(here also $A'_1=A_1$). The intuition
to understand this is that the contextual TTS merge too many states for the two
systems to remain differentiable. However we remark that here, the
first condition that we have required in Section~\ref{sec:nsc_problem_setting},
namely the global bisimulation between $\psi(\TTS{A'_1}[A'_2])$ and $\TTS{A_1}[A_2]$,
does not hold.

\subsubsection{Formalization}
Now we show that the conjunction of global and local bisimulations actually
gives the good definition.

\begin{defi}[Need for shared clocks]\label{def:nsc}
  Given $A_1\parallel A_2$ such that $A_1$ does not read the clocks of $A_2$,
  $A_2$ \emph{does not need to read the clocks of
    $A_1$} iff there exists an NTA $A'_1 \parallel A'_2$ without shared clocks
  (but with clock copies during synchronizations),
  using the same sets of local actions and a synchronization
  alphabet $\sync'$ related to the original one by a mapping $\psi: \sync'
  \times Q'_1 \rightarrow \sync \times Q_1$, and such that
  \begin{enumerate}
    \item $\psi(\TTS[Q_1']{A'_1}[A'_2])\wtb\TTS[Q_1]{A_1}[A_2]$ and
    \item $\psi(\TTS[Q'_1]{A'_1})\wtb\TTS[Q_1]{A_1}$ and
    \item $\psi(\TTS[A'_1]{A'_2})\wtb\TTS[A_1]{A_2}$.
  \end{enumerate}
\end{defi}

Notice that this does not mean that the clock constraints that read $X_1$ can
simply be removed from $A_2$ (see Fig.~\ref{fig:example1}).
\begin{lem}\label{lem:nsc}
  When there is no restriction in $\TTS[A_1]{A_2}$, any NTA $A'_1 \parallel
  A'_2$ which has no shared clocks and which satisfies items 2 and 3 of
  Definition~\ref{def:nsc}, also satisfies item 1.
\end{lem}
\proof
  When $\noRest[A_1]{A_2}$ holds, then by Lemma~\ref{lem:distrib},
  $\TTS[Q_1]{A_1}\x\TTS[A_1]{A_2}\stb\TTS[Q_1]{A_1}[A_2]$. So for any NTA
  $A'_1 \parallel A'_2$ satisfying items 2 and 3 of Definition~\ref{def:nsc},
  we have $\psi(\TTS[Q'_1]{A'_1})\x\psi(\TTS[A'_1]{A'_2}) \wtb
  \TTS[Q_1]{A_1}[A_2]$. It remains to show that
  $\psi(\TTS[Q'_1]{A'_1}[A'_2])\stb\psi(\TTS[Q'_1]{A'_1})\x\psi(\TTS[A'_1]{A'_2})$.
  Remark that applying $\psi$ to the labels before doing the product allows
  more synchronizations than applying $\psi$ on the TTS of the system since
  $\psi$ may merge different labels. We show that, in our case, the two
  resulting TTS are bisimilar anyway.

  For this, let $\R_1$ be a bisimulation relation between
  $\psi(\TTS[Q'_1]{A'_1})$ and $\TTS[Q_1]{A_1}$, and $\R_2$ be a
  bisimulation relation between $\psi(\TTS[A'_1]{A'_2})$ and
  $\TTS[A_1]{A_2}$. We will build inductively a bisimulation $\R$ between
  $\psi(\TTS[Q'_1]{A'_1}[A'_2])$ and
  $\psi(\TTS[Q'_1]{A'_1})\x\psi(\TTS[A'_1]{A'_2})$ such that for any $(q_1,
  q_2)$ and $(r_1, r_2)$ such that $(q_1, q_2) \R (r_1, r_2)$, there exists
  a state $s_1$ of $\TTS[Q_1]{A_1}$ and a state $s_2$ of
  $\TTS[A_1]{A_2}$ such that $q_1 \R_1 s_1$ and $r_1 \R_1 s_1$ and $q_2
  \R_2 s_2$ and $r_2 \R_2 s_2$.
  The inductive definition of $\R$ is as follows. The initial states (which
  are the same in both sides) are in relation; $\R$ is preserved by delays;
  $\R$ is preserved by playing local actions. The key is the treatment of
  synchronizations: when $(q_1, q_2) \R (r_1, r_2)$ and $q_1
  \xrightarrow{a_1} q'_1$ in $\TTS[Q_1]{A_1}$ and $q_2 \xrightarrow{a_2}
  q'_2$ in $\TTS[A_1]{A_2}$ with $\psi(a_1) = \psi(a_2) = a$, then the
  existence of the $s_1$ and $s_2$ mentioned earlier ensures that there
  exists a state $(r'_1, r'_2)$ in $\psi(\TTS[Q'_1]{A'_1}[A'_2])$ such that
  $(r_1, r_2) \xrightarrow{a} (r'_1, r'_2)$, and we set $(q'_1, q'_2) \R
  (r'_1, r'_2)$ for any such $(r'_1, r'_2)$.\qed

\subsubsection{A Criterion to Decide the Need for Shared Clocks}
We are now ready to give a criterion to decide whether shared clocks are necessary.
\begin{thm}\label{thm:nsc}
  When there is no restriction in $\TTS[A_1]{A_2}$ holds, $A_2$ does not
  need to read the clocks of $A_1$. When $A_2$ is deterministic, this
  condition becomes necessary.
\end{thm}

\proof[Proof of Theorem~\ref{thm:nsc}, necessary condition when $A_2$ is deterministic]
  Like in the proof of Lemma~\ref{lem:nsc}, we show that for any NTA
  $A'_1 \parallel A'_2$ satisfying items 2 and 3 of Definition~\ref{def:nsc},
  $\psi(\TTS[Q'_1]{A'_1}[A'_2]) \wtb \TTS[Q_1]{A_1}\x\TTS[A_1]{A_2}$. But, by
  Lemma~\ref{lem:distrib}, when $A_2$ is deterministic and $\TTS[A_1]{A_2}$
  has restrictions, $\TTS[Q_1]{A_1}\x\TTS[A_1]{A_2}$ is not timed bisimilar to
  $\TTS[Q_1]{A_1}[A_2]$ (not even weakly timed bisimilar since there are no
  $\eps$-transitions). Hence any NTA $A'_1 \parallel A'_2$ satisfying items
  2 and 3 of Definition~\ref{def:nsc}, does not satisfy item 1.\qed

We remark from the proof that when there is a restriction in $\TTS[A_1]{A_2}$,
even infinite $A'_1$ and $A'_2$ would not help. 
Next section will be devoted to the constructive proof of the direct part of
this theorem.

The counterexample in Fig.~\ref{fig:counterex} also works here to argue that the
conditions of Lemma~\ref{lem:nsc} and Theorem~\ref{thm:nsc} are not necessary
when $A_2$ is not deterministic. Indeed $A'_2$ with only one unguarded edge
labeled by $a$ and $A'_1=A_1$ satisfy the three items of
Definition~\ref{def:nsc} but there is a restriction in $\TTS[A_1]{A_2}$.

\section{Constructing a Network of Timed Automata without Shared Clocks}\label{sec:construction}
This section is dedicated to proving Theorem~\ref{thm:nsc} by constructing
suitable $A'_1$ and $A'_2$.
For simplicity, we assume that in $A_2$, the guards on the synchronizations do
not read $X_1$. Otherwise, the constraints that read $X_1$ could be moved
into the corresponding edges in $A_1$, with the intuition that, for a
synchronization, each automaton can check the constraints about its own clocks.

\subsection{Construction}\label{subsec:simple}
First, our $A'_1$ is obtained from $A_1$ by replacing all the labels
$a\in\sync$ on the synchronization edges of $A_1$ by
$(a,\ell_1)\in\sync\times L_1$, where $\ell_1$ is the output location of the
edge. Therefore the synchronization alphabet between $A'_1$ and $A'_2$ will
be $\sync'=\sync\times L_1$, which allows $A'_1$ to transmit its location
after each synchronization.

Then, the idea is to build $A'_2$ as a product $A_{1,2}\x A_{2,\mi{mod}}$
($\x$ denotes the product of TA as it is usually defined~\cite{AD94}),
where $A_{2,\mi{mod}}$ plays the role of $A_2$ and $A_{1,2}$ acts as a
local copy of $A'_1$, from which $A_{2,\mi{mod}}$ reads clocks instead of
reading those of $A'_1$.
For this, as long as the automata do not synchronize, $A_{1,2}$ will evolve,
simulating a run of $A'_1$ that is compatible with what $A'_2$ knows about
$A'_1$.
And, as soon as $A'_1$ synchronizes with $A'_2$, $A'_2$ updates
$A_{1,2}$ to the actual state of $A'_1$.
If the clocks of $A_{1,2}$ always give the same truth value to the guards and
invariants of $A_{2,\mi{mod}}$ than the actual value of the clocks of $A'_1$, then our
construction behaves like $A_1 \parallel A_2$. To check that this is the case,
we equip $A'_2$ with an error location, $\sad$, and edges that lead to it if there is a
contradiction between the values of the clocks of $A'_1$ and the values of the
clocks of $A_{1,2}$. The guards of these edges are the only cases where
$A'_2$ reads clocks of $A'_1$. Therefore, if $\sad$ is not
reachable, they can be removed so that $A'_2$ does not read the clocks of
$A'_1$.
More precisely, a contradiction happens when $A_{2,\mi{mod}}$ is in a given
location and the guard of an outgoing edge is true according to $A_{1,2}$ and
false according to $A'_1$, or vice versa, or when the invariant of the current
location is false according to $A'_1$ (whereas it is true according to
$A_{1,2}$, since $A_{2,\mi{mod}}$ reads the clocks of $A_{1,2}$).

Namely, $\S_\mi{mod}=A'_1\parallel (A_{1,2}\x A_{2,\mi{mod}})$
where $A_{1,2}$ and $A_{2,\mi{mod}}$ are defined as follows.
$A_{1,2}=(L_1,\ell_1^{0},X_1',\sync'\cup\{\eps\},E'_1,\inv_1')$, where
\begin{itemize}
  \item each clock $x'\in X'_1$ is associated
  with a clock $c(x')=x\in X_1$ ($c$ is a bijection from $X'_1$ to
  $X_1$).
  For any clock constraint $\gamma$, 
  $\gamma'$ denotes the clock constraint where any clock $x$ of $X_1$
  is substituted by $x'$ of $X'_1$.
  \item $\forall \ell\in L_1, \inv_1'(\ell)=\inv_1(\ell)'$
  \item $E'_1
  \begin{array}[t]{ll}
    =&\{\ell_1\xrightarrow{g',\eps,r'}\ell_2\mid
    \exists a\in\Sigma_1\setminus\Sigma_{2}^{\not\eps}:
    \ell_1\xrightarrow{g,a,c(r')}\ell_2\in E_1\}
    \\&\hfill
    \mbox{\it(simulate local actions of $A_1$)}\\
    \cup&\{\ell\xrightarrow{\true,(a,\ell_2),c}\ell_2\mid
    \ell \in L_1 \land
    a \in\sync \land
    \exists\ell_1\xrightarrow{g,a,r}\ell_2\in E_1\}
    \\& \qquad\qquad\qquad\hfill
    \mbox{\it(update the state of $A_{1,2}$ at each synchronization
      with $A_1$)}
  \end{array}$\\ 
  where 
  $c$ denotes the assignment of any clock $x'\in X'_1$
  with the value of its associated clock $c(x')=x\in X_1$ (written $x':=x$
  in Fig.~\ref{fig:example2}).
\end{itemize}
$A_{2,\mi{mod}}=(L_2\cup\{\sad\},\ell_2^{0},X_2\cup X'_1\cup X_1,
(\Sigma_2\setminus\Sigma_1)\cup\sync',E_2',\inv_2')$, where
\begin{itemize}
  \item $\forall\ell\in L_2,\inv_2'(\ell)=\inv_2(\ell)'$ and $\inv_2'(\sad)=\true$,
  \item $E'_2\begin{array}[t]{ll}
  =&\{\ell_1\xrightarrow{g',a,r}\ell_2
  \mid\ell_1\xrightarrow{g,a,r}\ell_2\in E_2 \land a\notin\sync\}
  \\
  \cup&\{\ell_1\xrightarrow{g,(a,\ell),r}\ell_2
  \mid\ell_1\xrightarrow{g,a,r}\ell_2\in E_2\land a \in\sync \land \ell \in L_1\}
  \\
  \cup&\{\ell\xrightarrow{\neg\inv_2(\ell),\eps,\emptyset}\sad
  \mid\ell\in L_2\}
  \\
  \cup&\{\ell\xrightarrow{g'\land\neg g,\eps,\emptyset}\sad
  \mid\ell\xrightarrow{g,a,r}\ell'\in E_2 \land a\notin\sync\}
  \\
  \cup&\{\ell\xrightarrow{\neg g'\land g,\eps,\emptyset}\sad
  \mid\ell\xrightarrow{g,a,r}\ell'\in E_2 \land a\notin\sync\}.
  \end{array}$
\end{itemize} 

\noindent For the example of Fig.~\ref{fig:example1}, $A_{1,2}$ and $A_{2,\mi{mod}}$
are pictured in Fig.~\ref{fig:example2}.
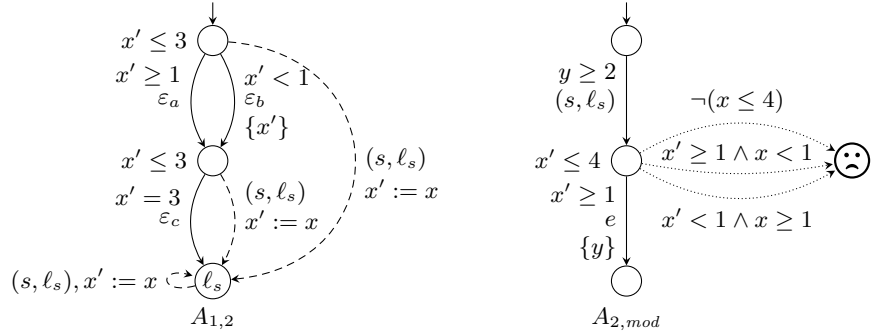
\begin{figure}[t]
  \centering
  \def\b{3.7}
\def\c{1.6cm}
\begin{tikzpicture}[node distance = \c, initial where = above]
  \node[state, initial] (q_0) [label=left:$x'\leq3$] {};
  \node[state] (q_1) [below of=q_0]
          [label=left:$x'\leq3$] {};
  \node[none] (a_1) at (0,-\b) {$A_{1,2}$};

  \node[state] (q_4) [below of=q_1] {$\ell_s$};
  \node[state, initial] (q_2) at (5.5,0) {};
  \node[state] (q_3) [below of=q_2] [label=left:$x'\leq4$] {};
  \node[state] (q_5) [below of=q_3] {};
  \node[sad] (q_6) at (8.5,-\c) {\huge\sad};
  \node[none] (a_2) at (5.5,-\b) {$A_{2,\mi{mod}}$};

  \path[->] (q_0) edge [bend right] node [swap, pos=0.45] {$x'\geq1$}
                                    node [swap] {$\eps_a$} (q_1)
            (q_1) edge [bend right,swap] node [pos=0.45] {$x'=3$}
                       node {$\eps_c$} (q_4)
            (q_0) edge [bend left] node [pos=0.45]{$x'<1$}
                                    node {$\eps_b$}
                                    node [pos=0.55] {$\{x'\}$} (q_1)
            (q_2) edge node [pos=0.2,swap] {$y\geq2$}
                       node [swap] {$(s,\ell_s)$} (q_3)
            (q_3) edge node [pos=0.2,swap] {$x'\geq1$}
                       node [swap]{$e$}
                       node [pos=0.8,swap] {$\{y\}$} (q_5);
  \path[->,densely dotted] (q_3) edge [bend left] node [above] {$\neg(x\leq4)$}(q_6)
                  edge [bend right=10] node [above]
                    {$x'\geq1\land x<1$}(q_6)
                  edge [bend right=30] node [below]
                    {$x'<1\land x\geq1$}(q_6);
  \path[->,densely dashed](q_1) edge [bend left]
                       node [pos=0.45] {$(s,\ell_s)$}
                       node [pos=0.5] {$x':=x$} (q_4)
            (q_0) edge [bend left=85, looseness=1.8]
                       node [pos=0.5] {$(s,\ell_s)$}
                       node [pos=0.53] {$x':=x$}(q_4)
            (q_4) edge [loop left] node {$(s,\ell_s),x':=x$}(q_4);
\end{tikzpicture}
  \caption{$A_{1,2}$ and $A_{2,\mi{mod}}$ for the example of
  Fig.~\ref{fig:example1} \label{fig:example2}. We represent by dotted arcs the
  edges leading to the error state, and by dashed arcs those used during
  synchronizations to reset $A_{1,2}$ to the actual state of $A_1$.}
\end{figure}


We now prove the
correspondence between a state of $\S_\mi{mod}$ and two states of
$\TTS{A_1}[A_2]$ that are merged into the same state of $\TTS[A_1]{A_2}$.
This is stated in the following proposition.
A state of $\S_\mi{mod}$ is denoted as $(s_1,s_{1,2},s_2)=
\big((\ell_1,v_{|X_1}),(\ell_{1,2},v_{|X'_1}),
(\ell_2,v_{|X_2\setminus X_1})\big)$. 
For a given state of $A_{1,2}$, $s_{1,2}=(\ell_{1,2},v_{|X'_1})$, we denote
by $s'_{1,2}$ the state $(\ell_{1,2},v')$, where $v':X_1\to\Reals$ is defined as:
for any $x\in X_1$, $v'(x)=v(x')$ (\ie $s'_{1,2}$ is a state of $A_1$).
Reciprocally, for a given
state of $A_1$, $s'_{1,2}=(\ell_{1,2},v')$,
$s_{1,2}$ denotes the state $(\ell_{1,2},v)$, where $v:X'_1\to\Reals$ is defined as:
for any  $x'\in X'_1$, $v(x')=v'(x)$.
\begin{prop}\label{prop:states}
  Let $(s_1,s_{1,2},s_2)$ 
  be a state of $\S_\mi{mod}$. 
  If along one path that leads to $(s_1,s_{1,2},s_2)$ no edge leading to
  $\sad$ is enabled, then there exists $S_1$ such that $(S_1,s_2)$
  is a reachable state of $\TTS[A_1]{A_2}$ and $s_1$ and $s'_{1,2}$ are
  both in $S_1$.

  Conversely, let $(S_1,s_2)$ be a reachable state of $\TTS[A_1]{A_2}$, and
  $s_1$ and $s'_{1,2}$ be some states in $S_1$.
  Then $(s_1,s_{1,2},s_2)$ is a state of $\S_\mi{mod}$.
\end{prop}

\proof
  Let $(s_1,s_{1,2},s_2)$ be a reachable state of $\S_\mi{mod}$, such that
  there is a path $\rho$ from the initial state $(s_1^0,s_{1,2}^0,s_2^0)$
  to $(s_1,s_{1,2},s_2)$ that does not enable any edges leading to $\sad$
  (except maybe from $(s_1,s_{1,2},s_2)$).
  We give a recursive proof. First, for the initial state
  $(s_1^0,s_{1,2}^0,s_2^0)$ of $\S_\mi{mod}$, $s_1^0$ and
  $s_{1,2}^{0\prime}$ are both in $S_1^0$ such that $(S_1^0,s_2^0)$ is the
  initial state of $\TTS[A_1]{A_2}$. Now, assume this is true for some
  $(p_1,p_{1,2},p_2)$ visited along $\rho$. That is, there exists $P_1$
  such that $(P_1,p_2)$ is reachable and $p_1,p'_{1,2}\in P_1$.
  Then, the next state $s'$ visited along $\rho$ is reached
  after one of the
  following steps:
  \begin{itemize}
    \item local action in $A'_1$: $s'=(q_1,p_{1,2},p_2)$ such that
    $q_1\in\UR(p_1)\subseteq P_1$,
    \item local action in $A_{1,2}$: $s'=(p_1,q_{1,2},p_2)$ such that
    $q'_{1,2}\in\UR(p'_{1,2})\subseteq P_1$,
    \item local action in $A_2$: $s'=(p_1,p_{1,2},q_2)$ such that
    there exists $S'_1$ such that $(S'_1,q_2)$ is reachable from $(P_1,q_2)$
    by the same action, and, since no edge leading to $\sad$ is enabled,
    both $(p_1,p_2)$ and $(p'_{1,2},p_2)$ enable this step in
    $\TTS{A_1}[A_2]$. Therefore, $p_1,p'_{1,2}\in S'_1$.
    \item synchronization: $s'=(q_1,q_{1,2},q_2)$ such that there exists
    $S'_1=\UR(q_1)$ such that $(S'_1,q_2)$ is reachable from $(P_1,q_2)$
    by the same action, and $q_1=q'_{1,2}\in S'_1$.
  \end{itemize}
  By recursion, $(s_1,s_{1,2},s_2)$ also satisfies the property, that is,
  there exists $S_1$ such that $(S_1,s_2)$ is reachable and
  $s_1,s'_{1,2}\in S_1$.

  Conversely, let denote by $P(S_1,s_2)$ the fact that for any reachable
  state $(S_1,s_2)$ of $\TTS[A_1]{A_2}$, for any states
  $s_1,s'_{1,2}\in S_1$, $(s_1,s_{1,2},s_2)$ is a reachable state of
  $\S_\mi{mod}$.
  First, for any $s_1,s'_{1,2}\in S_1^0=\UR(s_1^0)$, $(s_1,s_{1,2},s_2^0)$
  is a reachable state, because by construction, $A_{1,2}$ can only mimic
  (as long as there is no synchronization) one possible behavior of $A_1$
  to reach $s_{1,2}$ from $s_1^0$, therefore $P(S_1^0,s_2^0)$
  holds.
  Assume that for some reachable state $(S_1,s_2)$ $P(S_1,s_2)$ holds.
  Then any state reachable in one step from $(S_1,s_2)$ is reached by one of
  the following steps.
  \begin{itemize}
  \item If for some $a\in \Sigma_2\setminus\sync$,
    $(S_1,s_2)\xrightarrow{a}(S'_1,s'_2)$, then for any
    \mbox{$s_1,s'_{1,2}\in S'_1\subseteq S_1$},
    $(s_1,s'_{1,2},s_2)\xrightarrow{a}(s_1,s'_{1,2},s'_2)$,
    \ie $P(S'_1,s'_2)$ holds.
  \item If for some $(a,s'_1)\in \sync\times Q_1$,
    $(S_1,s_2)\xrightarrow{a,s'_1}(S'_1,s'_2)$, then $S'_1=\UR(s'_1)$, and
    for any \mbox{$s_1,s'_{1,2}\in S'_1$}, $(s_1,s_{1,2},s'_2)$ can be
    reached from some $(p_1,p_{1,2},s_2)$ such that \mbox{$p_1,p'_{1,2}\in S_1$}.
    Indeed, in $\S_\mi{mod}$, synchronization $((a,\ell'_1),s'_1)$ resets
    $A_{1,2}$ in the same state as $A_1$ and then
    $A_1$ performs some local actions while $A_{1,2}$ also performs some
    local actions mimicking one possible behavior of $A_1$ (that is why
    $s'_{1,2}\in S'_1$).
    Hence $P(S'_1,s'_2)$ holds.
  \item If for some $d\in \Reals$,
    $(S_1,s_2)\xrightarrow{d}(S'_1,s'_2)$, then we use the same reasoning as
    for a synchronization. Since $A_{1,2}$ is built
    so that it mimics any possible behavior of $A_1$ between synchronizations,
    any state $s'_{1,2}\in S'_1$ reachable by $A_1$ during this delay
    corresponds to a state $s_{1,2}$ reachable by $A_{1,2}$.
    Hence $P(S'_1,s'_2)$ also holds.
  \end{itemize}
  By recursion, $P(S_1,s_2)$ holds for any reachable state $(S_1,s_2)$.\qed

Lastly, the following lemma will be used to prove the direct part of
Theorem~\ref{thm:nsc}.

\begin{lem}\label{lem:sad_equiv_read}
  $\sad$ is reachable in $\S_\mi{mod}$ iff there is a restriction
  in $\TTS[A_1]{A_2}$.
\end{lem}
\proof
  Assume $\sad$ is not reachable in $\S_\mi{mod}$. From
  Proposition~\ref{prop:states}, we know that for any state $(S_1,s_2)$ of
  $\TTS[A_1]{A_2}$, for any $s_1$, $s'_{1,2}$ in $S_1$, there is a
  corresponding state $s=\big((\ell_1,v_{|X_1}),(\ell_{1,2},v_{|X'_1}),
  (\ell_2,v_{|X_2\setminus X_1})\big)=(s_1,s_{1,2},s_2)$ of $\S_\mi{mod}$. Moreover, for
  any such $s$, if there is an outgoing edge towards $\sad$ from $\ell_2$,
  then this edge is never enabled. That is, for any time constraint $\gamma$
  read in $\ell_2$ in the original system $\S$ (invariant of $\ell_2$ or
  guard of an outgoing edge with a local action),
  $v_{|X_2\cup X_1}\models \gamma\iff
  v_{|(X_2\setminus X_1)\cup X'_1}\models \gamma'$.
  Hence for any enabled step from $(S_1,s_2)$, $s_1$ and $s'_{1,2}$ are in the
  same restriction. Therefore, $\noRest[A_1]{A_2}$.

  Assume $\sad$ is reachable in $\S_\mi{mod}$. From
  Proposition~\ref{prop:states}, we know that for any state
  $s=\big((\ell_1,v_{|X_1}),(\ell_{1,2},v_{|X'_1}),(\ell_2,v_{|X_2\setminus X_1})\big)
  =(s_1,s_{1,2},s_2)$ of $\S_\mi{mod}$, reached after a path that does not
  enable edges leading to $\sad$ (except maybe from this last state), there
  is a corresponding state $(S_1,s_2)$ of $\TTS[A_1]{A_2}$ such that
  $s_1$ and $s'_{1,2}$ are both in $S_1$.
  If $\sad$ can be reached, then consider a path that reaches $\sad$
  and such that no edge leading to $\sad$ was enabled before along the path.
  The last state $s$ of $\S_\mi{mod}$ visited before $\sad$ is such that
  for some time constraint $\gamma$ evaluated at $s$ from $\ell_2$,
  $v_{|X_2\cup X_1}\models \gamma$ and
  $v_{|(X_2\setminus X_1)\cup X'_1}\not\models \gamma'$ (or conversely).
  Therefore, a local action or local delay is possible from $(s_1,s_2)$ and
  not from $(s'_{1,2},s_2)$.
  Hence $(S_1,s_2)$ is a state with a restriction.\qed

%

We now give a first simple case for which Theorem~\ref{thm:nsc} can be proved
easily. We say that $A_1$ has no urgent synchronization if for any location,
when the invariant reaches its limit, a local action is enabled.
Under this assumption, we can show that
$A'_2=A_{1,2}\x A'_{2,\mi{mod}}$, where $A'_{2,\mi{mod}}$ is
$A_{2,\mi{mod}}$ without location $\sad$ (that is never reached according to
Lemma~\ref{lem:sad_equiv_read}) and its ingoing edges, is suitable.
Indeed, we can show that $A'_2$ does not read $X_1$ and is such that
$\psi(\TTS[A'_1]{A'_2})\wtb\TTS[A_1]{A_2}$, where for any
$((a,\ell_1),s_1)\in \sync'\times Q'_1$, $\psi(((a,\ell_1),s_1))=(a,s_1)$.
Obviously, item 2 of Definition~\ref{def:nsc} holds, and Lemma~\ref{lem:nsc}
says that item 1 also holds.

When $A_1$ has urgent synchronizations, this construction allows one to check
the absence of restriction in $\TTS[A_1]{A_2}$, but it does not give directly
a suitable $A'_2$. We define the construction of $A'_2$ for
the general case in Subsection~\ref{subsec:general}.


\proof[Proof of Theorem~\ref{thm:nsc},
direct part, when no urgent synchronization in $A_1$]\label{proof}\hfill

  \noindent Assume $\noRest[A_1]{A_2}$.
  We consider
  $A'_2=A_{1,2}\x A'_{2,\mi{mod}}$ where $A'_{2,\mi{mod}}$ is
  $A_{2,\mi{mod}}$ without $\sad$ (that is never reached according to
  Lemma~\ref{lem:sad_equiv_read}) and its ingoing edges. Therefore,
  $A'_{2,\mi{mod}}$ does not read $X_1$ and neither does
  $A'_2=A_{1,2}\x A'_{2,\mi{mod}}$. Below we show that $A'_2$ is a suitable
  candidate because $\psi(\TTS[A'_1]{A'_2})\wtb\TTS[A_1]{A_2}$
  ($\psi(\TTS[Q'_1]{A'_1})\wtb\TTS[Q_1]{A_1}$ obviously holds).

  Let $\R$ be the relation such that for any reachable state $(S_1,s_2)$ of
  $\TTS[A_1]{A_2}$, and any reachable state $(S'_1,s'_2)$ of $\psi(\TTS[A'_1]{A'_2})$,
  \[(S_1,s_2)\R(S'_1,s'_2) \iffdef\left\{
  \begin{array}{l}
    s_2=(\ell_2, v_2)\text{ and }s'_2=((\ell_{1,2},\ell_2),v'_2)\text{ s.t.}\\
    \forall x\in X_2\setminus X_1, v_2(x)=v'_2(x)\\
    S_1=S'_1\\
  \end{array}
  \right.\]
  \ie $A_2$ and $A'_{2,\mi{mod}}$ are both in $\ell_2$ and their local
  clocks have the same value, and $A_1$ and $A'_1$ are in
  indistinguishable states (states merged in a same contextual state $S_1$).
  Obviously, the initial states, $(S_1^0,s_2^0)$ and
  $(S_1^0,s_2^{0\prime})$, are $\R$-related.
  Since there is no marked state in $\TTS[A_1]{A_2}$ (resp.\ in
  $\TTS[A'_1]{A'_2}$), for any state $s=(S_1,s_2)$ (resp.\
  $s'=(S'_1,s'_2)$) of this TTS, all time constraints read by automaton 2 in
  $\ell_2$ (invariant of $\ell_2$ and guards of the outgoing edges) have the
  same truth value for all the states $(s_1,s_2)$ such that $s_1\in S_1$
  (resp.\ $s_1\in S'_1$).
  In the sequel, we say that valuation $V$ of $s$ (resp.\ $V'$ of $s'$) satisfies
  constraint $g$, when the valuations of all states $(s_1,s_2)$ in $s$
  (resp.\ in $s'$) satisfy $g$.
  Assume now that for some reachable states $(S_1,s_2)$ and $(S'_1,s'_2)$,
  $(S_1,s_2)\R(S'_1,s'_2)$.

  \subsubsection*{Local Action}
  If $a\in\Sigma_2\setminus\Sigma_1$ is enabled from $(S_1,s_2)$, then,
  there is an associated edge in $A_2$, $\ell_2\xrightarrow{g,a,r}p_2$ such that
  guard $g$ is satisfied by $V$.
  Let $g'$ be the guard on the corresponding outgoing edge
  $(\ell_{1,2},\ell_2)\xrightarrow{g',a,r}(\ell_{1,2},p_2)$ in $A'_2$.
  $g$ uses clocks in $X_2$, and by construction,
  $g'$ has the same form but with clocks in $(X_2\setminus X_1)\uplus X'_1$.
  $(S_1,s_2)\R(S'_1,s'_2)$ says that $v_2$ and $v'_2$ coincide on $X_2\setminus X_1$, and since
  $\sad$ is never reached in $\S_\mi{mod}$, $V$ satisfies the constraints of $g$
  on $X_1$ iff $V'$ satisfies the constraints of $g'$ on $X_1'$.
  That is, $V\models g\iff V'\models g'$. Therefore $A'_2$ can also perform
  $a$ from $(S_1,s'_2)$ and the states reached in both systems
  are $\R$-related: $(S_1,q_2)\R(S_1,q'_2)$, because $q_2=(p_2,v_2[r])$ and
  $q'_2=((\ell_{1,2},p_2),v'_2[r])$.
  This also holds reciprocally.

  \subsubsection*{Synchronization} Assume for some $(a,s'_1)\in\sync\times Q_1$,
  $(S_1,s_2)\xrightarrow{a,s'_1}(S_1',q_2)$.
  That is, there is an edge
  $\ell_2\xrightarrow{g_2,a,r_2} p_2$  in $A_2$ such that $v_2\models g_2$
  and $q_2=(p_2,v_2[r_2])$ and,
  for some $(\ell_1,v_1)\in S_1$, an edge $\ell_1\xrightarrow{g_1,a,r_1} p_1$
  in $A_1$ such that $v_1\models g_1$ and $s'_1=(p_1,v_1[r_1])\in S'_1$.
  Hence, synchronization $((a,p_1),s'_1)$ is also enabled from state $(S_1,s'_2)$
  because $A_{2,\mi{mod}}$ is in the same location as $A_2$, and has the same
  clock values over $X_2\setminus X_1$, and $A'_1$ is also in some state
  of $S_1$, therefore, there is also the same state $(\ell_1,v_1)\in S_1$
  which enables $(a,p_1)$.
  We do not consider $A_{1,2}$ because it is always ready to
  synchronize.
  Moreover, the state reached in $\psi(\TTS[A'_1]{A'_2})$ after this synchronization is
  $(S'_1,q'_2)$ such that $(S'_1,q_2)\R(S'_1,q'_2)$, because $q_2=(p_2,v_2[r_2])$
  and $q'_2=\big((p_{1,2},p_2),(v'_2[r_2])[c]\big)$ where $c$ denotes the copy of the
  clocks of $X_1$ into their associated clocks of $X'_1$ and therefore $c$ modifies
  only clocks that we do not consider in relation $\R$, and $r_2\subseteq C_2\subseteq(X_2\setminus X_1)$
  resets the same clocks in both systems.
  And reciprocally.

  \subsubsection*{Local Delay}
  Assume for some $d\in\Reals$, $(S_1,s_2)\xrightarrow{d}(S'_1,q_2)$.
  Then, \mbox{$V+d\models\inv_2(\ell_2)$}, and since $\sad$ is never reached
  in $\S_\mi{mod}$, $V+d\models\inv_2(\ell_2)\iff V'+d\models\inv'_2(\ell_2)$.
  That is, the same delay is enabled from $(S_1,s'_2)$ while $A_{1,2}$ may
  perform some local steps:
  $(S_1,s'_2){(\xrightarrow{g_0,\eps,r_0})}^*\xrightarrow{d_0}
  {(\xrightarrow{g_n,\eps,r_n})}^*\dots\xrightarrow{d_n}(S''_1,q'_2)$,
  where $\sum_{i=0}^nd_i=d$, $g_i$ is a guard over $X'_1$ and $r_i$ is a reset
  included in $X'_1$.
  This works because we assumed that $A_1$ has no urgent synchronization
  (and so does $A'_1$).
  Therefore, $A_{1,2}$ cannot force a synchronization.

  Reciprocally, if we can perform a delay $d$ from $(S_1,s'_2)$, then
  $V'+d\models\inv'_2(\ell_2)\land\inv'_1(\ell_{1,2})$.
  And since
  $V+d\models\inv_2(\ell_2)\iff V'+d\models\inv'_2(\ell_2)$,
  we can perform the same delay from $(S_1,s_2)$.

  Moreover, we reach equivalent states in both systems. Indeed, $A_2$ and
  $A'_{2,\mi{mod}}$ stay in the same location, the clocks in $X_2\setminus X_1$
  increase their value by $d$, 
  and the set of states of $A_1$ and $A'_1$ becomes
  $S'_1=S''_1=\{s'_1\mid \exists s_1\in S_1,\rho\in \paths{\Sigma_1\setminus\Sigma_{2}^{\not\eps},d}:
  (s_1,s_2)\xRightarrow{\rho}(s'_1,q_2)\}$.

  Therefore, $\R$ is a weak timed bisimulation and
  $\psi(\TTS[A'_1]{A'_2})\wtb\TTS[A_1]{A_2}$.
  Lastly, by Lemma~\ref{lem:nsc}, $\psi(\TTS[Q'_1]{A'_1}[A'_2])\wtb\TTS[Q_1]{A_1}[A_2]$
  also, and $A_2$ does not need to read $X_1$.\qed

In the example of Fig.~\ref{fig:example1}, $\sad$ is not reachable in
$\S_{\mi{mod}}$ (see Fig.~\ref{fig:example2}), therefore $A_2$ does not need
to read $X_1$.
For an example where $\sad$ is reachable, consider the same example with an
additional edge $\xrightarrow{\true,f,\{x\}}$ from the end location of $A_1$
to a new location. Location $\sad$ can now be reached in $\S_{\mi{mod}}$,
for example consider a run where $s$ is performed at time 2 leading to a state
where $v(x)=2$ and $v(x')=2$, and then $A_1$ immediately performs $f$
and resets $x$, leading to a state where the valuation $v'$ is such that
$v'(x)=0$ and $v'(x')=2$, and satisfies guard $x'\geq1\land x<1$ in
$\S_{\mi{mod}}$. Therefore, with this additional edge in $A_1$, $A_2$
needs to read $X_1$. Indeed, without this edge, $A_2$ knows that $A_1$
cannot modify $x$ after the synchronization, but with this edge, $A_2$ does
not know whether $A_1$ has performed $f$ and reset $x$, while this may
change the truth value of its guard $x\geq 1$.

\subsection{Complexity}
\subsubsection*{\textsf{PSPACE}-hardness}
The reachability problem for timed automata is known to be
\textsf{PSPACE}-complete~\cite{AD90}. We will reduce this problem to our problem
of deciding whether $A_2$ needs to read the clocks of $A_1$.
Consider a timed automaton $A$ over alphabet $\Sigma$,
with some location $\ell$. Build the timed
automaton $A_2$ as $A$ augmented with two new locations $\ell'$ and $\ell''$
and two edges, $\ell\xrightarrow{\true,\eps,\emptyset}\ell'$ and
$\ell'\xrightarrow{x=1,a,\emptyset}\ell''$, where $x$ is a fresh clock, and
$a$ is some action in $\Sigma$.
Let $A_1$ be the one of Fig.~\ref{fig:counterex} with an action $b\notin\Sigma$.
Then, $\ell$ is reachable in $A$ iff $A_2$ needs to
read $x$ which belongs to $A_1$. Therefore the problem of deciding whether
$A_2$ needs to read the clocks of $A_1$ is also \textsf{PSPACE}-hard.

\subsubsection*{\textsf{PSPACE}-membership}
Moreover, we can show that when $A_2$ is deterministic, our problem is in
\textsf{PSPACE}. Indeed, by Theorem~\ref{thm:nsc} and
Lemma~\ref{lem:sad_equiv_read}, $\sad$ is not reachable iff
$\noRest[A_1]{A_2}$ iff $A_2$ does not need to read the clocks of $A_1$.
Since the size of the modified system on which we check the reachability of
$\sad$ is polynomial in the size of the original system, our problem is in
\textsf{PSPACE}.

\subsection{Dealing with Urgent Synchronizations}
\label{subsec:general}

If we use exactly the same construction as before and allow urgent
synchronizations, the following problem may occur. Remind that $A_{1,2}$
simulates a possible run of $A'_1$ while $A'_1$ plays its actual run. There is
no reason why the two runs should coincide. Thus it may happen that the run
simulated by $A_{1,2}$ reaches a state where the invariant expires and only a
synchronization is possible. Then $A'_2$ is expecting a synchronization with
$A'_1$, but it is possible that the actual $A'_1$ has not reached a state
that enables this synchronization. Intuitively,
$A'_2$ should then realize that the simulated run cannot be the actual one and
try another run compatible with the absence of synchronization.
%
%

In fact, between two synchronizations,
$A_{1,2}$, the local copy of $A_1$, can be constructed to simulate
only one fixed run of $A_1$, instead of being able to simulate all its runs.
If this run is well chosen, then the situation described above never happens,
and we can use a construction similar to the one above, on which we can prove that
if $\sad$ is not reachable, then any run of $A_1$ is compatible with the
fixed run of $A_{1,2}$, and $A_2$ can avoid reading the clocks of $A_1$.

Therefore, the idea of the construction is to force $A_{1,2}$ to simulate one
of the runs of $A_1$ (from the state reached after
the last synchronization) that has maximal duration before it synchronizes again
with $A_{2,\mi{mod}}$ (or never synchronizes again if
possible). There may not be any
  such run if some time constraints are strict inequalities, but the idea can be
  adapted even to this case.
This choice of a run of $A_1$ is as valid as the others, and it
prevents the system from having to deal with the subtle situation that we described
above.
Below, we describe the construction of $A_{1,2}$ in two cases:
\begin{enumerate}
  \item After any synchronization there is a local run of maximal duration.
  \item It may happen that, after a synchronization, there is no
    run of maximal duration because of some strict time constraints.
\end{enumerate}

\subsubsection*{Case 1:
After any synchronization there is a local run of maximal duration}
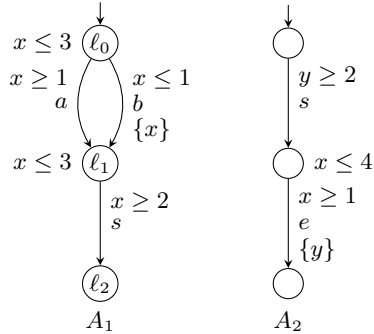
\begin{figure}[tp]
  \centering
  \def\b{3.7}
\begin{tikzpicture}[node distance = 1.6cm, initial where = above]
  \node[state, initial] (q_0) [label=left:$x\leq3$] {$\ell_0$};
  \node[state] (q_1) [below of=q_0]
          [label=left:$x\leq3$] {$\ell_1$};
  \node[none] (a_1) at (0,-\b) {$A_1$};
  \node[state] (q_4) [below of=q_1] {$\ell_2$};

  \node[state, initial] (q_2) at (2.5,0) {};
  \node[state] (q_3) [below of=q_2] [label=right:$x\leq4$] {};
  \node[state] (q_5) [below of=q_3] {};
  \node[none] (a_2) at (2.5,-\b) {$A_2$};

  \path[->] (q_0) edge [bend right] node [swap, pos=0.45] {$x\geq1$}
                                    node [swap] {$a$} (q_1)
            (q_1) edge node [pos=0.25] {$x\geq2$}
                       node {$s$} (q_4)
            (q_0) edge [bend left] node [pos=0.45]{$x\leq1$}
                                    node {$b$}
                                    node [pos=0.55] {$\{x\}$} (q_1);

  \path[->] (q_2) edge node [pos=0.2]{$y\geq2$} node {$s$} (q_3)
            (q_3) edge node [pos=0.2] {$x\geq1$}
                                   node {$e$}
                                   node [pos=0.8] {$\{y\}$}(q_5);
\end{tikzpicture}
  \caption{$A_1$ has an urgent synchronization.}\label{fig:urgent}
\end{figure}
Consider automaton $A_1$ in Fig.~\ref{fig:urgent}.
We can see that, for the urgent synchronization to happen as late as possible,
$A_{1,2}$ has to fire $b$ at time 1, so that it can then wait 3 time units
before synchronizing, although it is still able to synchronize at any time (we
add the same dashed edges as in Fig.~\ref{fig:example2}). Fig.~\ref{fig:result}
shows a timed automaton that achieves the desired behaviour for $A_{1,2}$
using a fresh clock $z$ to force the simulation of $b$ at time $1$.

\begin{figure}[t]
  \centering
  \def\b{3.7}
\def\c{1.6cm}
\begin{tikzpicture}[node distance = \c, initial where = above]
  \node[state, initial] (q_0) [label=left:$z\leq1$] {};
  \node[state] (q_1) [below of=q_0]
          [label=left:$x'\leq3$] {};
  \node[state] (q_4) [below of=q_1] {$\ell_2$};

  \path[->] (q_0) edge [swap]
                       node [pos=0.25]{$z=1$}
                       node {$\eps_b$}
                       node [pos=0.75] {$\{x',z\}$} (q_1);
  \path[->,densely dashed](q_1) edge [swap]
                       node [pos=0.35] {$(s,\ell_2)$}
                       node [pos=0.65] {$x':=x$} (q_4)
            (q_0) edge [bend left=85, looseness=1.8]
                       node [pos=0.5] {$(s,\ell_2)$}
                       node [pos=0.53] {$x':=x$}(q_4)
            (q_4) edge [loop left] node {$(s,\ell_2),x':=x$}(q_4);
\end{tikzpicture}
  \caption[$A_{1,2}$ associated with $A_1$ of Fig.~\ref{fig:urgent}]
  {$A_{1,2}$ associated with $A_1$ of Fig.~\ref{fig:urgent}.}\label{fig:result}
\end{figure}
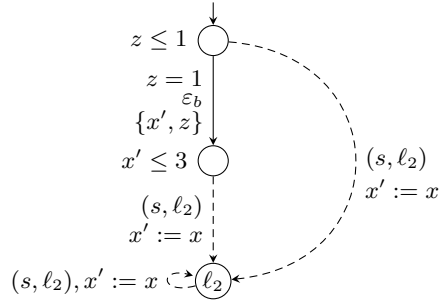

This can be generalized for any $A_1$. The idea is essentially to force
$A_{1,2}$ to follow the appropriate finite or ultimately periodic path in the
region automaton~\cite{AD94} of $A_1$.
The construction is described below and illustrated by
Fig.~\ref{fig:region}.

$A_{1,2}$ is now built over the region automaton~\cite{AD94}
of $A_1$. Transitions labeled by some $a\in\sync$ are treated separately
like in the original construction. The problem now is to constrain $A_{1,2}$
to take one of the most time consuming local runs after a synchronization.

The first step is to build the region automaton of $A_1$, and remove the
synchronizations. Then, from each state $s$ we compute the most time consuming
run and keep only the output arcs of $s$ that start a most time consuming run.

The computation of the most time consuming runs from $s$ is done as follows.
If one of the paths from $s$ has a loop, then there is an infinite run from
$s$ with local actions, and since we consider non-Zeno TA, time diverges and
this run is valid.
If no path from $s$ contains a loop, then
the paths from $s$ are finite and there is a finite number of such paths.
It is possible to compute, for each path, the supremum of the duration of the path:
just sum the maximal delays in each location (including the time spent in the
last location).

It remains to force, using a fresh clock, the longest stay in each state.

\begin{figure}[tp]
  \centering
  \def\b{1.8}
\begin{tikzpicture}[node distance = 1.6cm, initial where = above]
  \tikzstyle{state} = [rectangle, rounded corners,
                   draw,
                   inner sep      = 2pt,
                   text width     = 1cm,
                   minimum height = 0.8cm,
                   text centered]
  \node[state, initial] (q_0) {$\ell_0$ $x=0$};
  \node[state,text width = 1.5cm] (q_01) at (\b,0)   {$\ell_0$ \mbox{$0<x<1$}};
  \node[state] (q_02) at (2*\b,0) {$\ell_0$ $x=1$};
  \node[state,text width = 1.5cm] (q_03) at (3*\b,0) {$\ell_0$ \mbox{$1<x<2$}};
  \node[state] (q_04) at (4*\b,0) {$\ell_0$ $x=2$};
  \node[state,text width = 1.5cm] (q_05) at (5*\b,0) {$\ell_0$ \mbox{$2<x<3$}};
  \node[state] (q_06) at (6*\b,0) {$\ell_0$ $x=3$};

  \node[state] (q_1) [below of=q_0] {$\ell_1$ $x=0$};
  \node[state,text width = 1.5cm] (q_12) at (1*\b,-1.6) {$\ell_1$ \mbox{$0<x<1$}};
  \node[state] (q_13) at (2*\b,-1.6) {$\ell_1$ $x=1$};
  \node[state,text width = 1.5cm] (q_14) at (3*\b,-1.6) {$\ell_1$ \mbox{$1<x<2$}};
  \node[state] (q_15) at (4*\b,-1.6) {$\ell_1$ $x=2$};
  \node[state,text width = 1.5cm] (q_16) at (5*\b,-1.6) {$\ell_1$ \mbox{$2<x<3$}};
  \node[state] (q_17) at (6*\b,-1.6) {$\ell_1$ $x=3$};

  \node[state] (q_2) [below of=q_15] {$\ell_2$ $x=2$};
  \node[state,text width = 1.5cm] (q_22)  [below of=q_16] {$\ell_2$ \mbox{$2<x<3$}};
  \node[state] (q_23) [below of=q_17] {$\ell_2$ $x=3$};
  \node[state] (q_24) at (7*\b,-3.2) {$\ell_2$ $x>3$};

  \tikzstyle{every node}=[inner sep=1pt]
  \path[->] (q_0)  edge [dashed,swap] node {$b$} (q_1)
            (q_01) edge [dashed,swap] node {$b$} (q_1)
            (q_02) edge [swap] node {$b$} (q_1)
            (q_02) edge [dashed] node {$a$} (q_13)
            (q_03) edge [dashed] node {$a$} (q_14)
            (q_04) edge [dashed] node {$a$} (q_15)
            (q_05) edge [dashed] node {$a$} (q_16)
            (q_06) edge [dashed] node {$a$} (q_17)
            (q_15) edge [dotted] node {$s$} (q_2)
            (q_16) edge [dotted] node {$s$} (q_22)
            (q_17) edge [dotted] node {$s$} (q_23)
            (q_0)  edge node {$\eps$} (q_01)
            (q_01) edge node {$\eps$} (q_02)
            (q_02) edge node {$\eps$} (q_03)
            (q_03) edge node {$\eps$} (q_04)
            (q_04) edge node {$\eps$} (q_05)
            (q_05) edge node {$\eps$} (q_06)
            (q_1)  edge node {$\eps$} (q_12)
            (q_12) edge node {$\eps$} (q_13)
            (q_13) edge node {$\eps$} (q_14)
            (q_14) edge node {$\eps$} (q_15)
            (q_15) edge node {$\eps$} (q_16)
            (q_16) edge node {$\eps$} (q_17)
            (q_2)  edge node {$\eps$} (q_22)
            (q_22) edge node {$\eps$} (q_23)
            (q_23) edge node {$\eps$} (q_24);
\end{tikzpicture}
  \caption[Region automaton of $A_1$ of Fig.~\ref{fig:urgent}]
  {The region automaton of $A_1$ of Fig.~\ref{fig:urgent}. The dashed arcs
    indicate occurrences of internal actions of $A_1$ that will be removed in
    the construction of $A_{1, 2}$ in order to force a run of maximal duration
    between synchronizations. The occurrences of the synchronization $s$,
    represented by dotted arrows, are treated separately in the construction
    of $A_{1, 2}$.}
  \label{fig:region}
\end{figure}
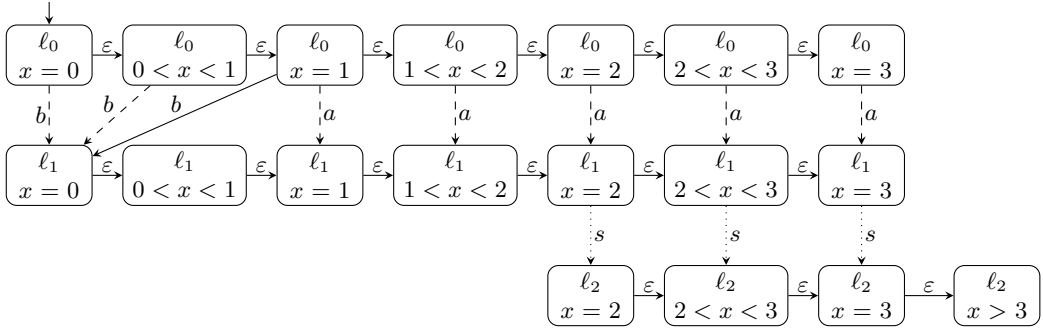

Lastly, we treat the synchronizations like in the construction of
Section~\ref{subsec:simple}: for each synchronizing edge in $A_1$, and each
corresponding output state in the region automaton,
we add synchronizing edges from all states of $A_{1,2}$, which reset the state
of $A_{1,2}$ to the actual state of $A_1$. These edges are labeled by
``$\gamma(R),(a,\ell_1),c$'', where $\gamma(R)$ is the constraint that
describes the region $R$ associated with the target state, $a$ is the
synchronization label in $A_1$, $\ell_1$ is the output location of the
synchronization in $A_1$, and $c$ is the assignment of clock values.

\subsubsection*{Definition of $A_{1,2}$}
Assume $(S,s_0,E)$ is a structure that
stores the region automaton of $A_1$, without the synchronization edges,
and with only the edges that are in the most time consuming paths computed as explained earlier.
That is, $S$ (resp.\ $s_0$) is the set of states (resp.\ the initial state)
of the region automaton of $A_1$, and
$E\subseteq S\times(\N\times E_1)\times S$
stores edges in the form \mbox{$s\xrightarrow{d,e}s'$}
where $d$ is the delay that has to be performed in $\ell(s)$, the location
associated with state $s$, before performing edge $e$ labeled by some
action in $\Sigma_1\setminus\sync$.
Then, $A_{1,2}=(S,s_0,X_1\cup C'_1\cup\{z\},\sync'\cup\{\eps\},E'_1,\inv'_1)$ where
\begin{itemize}
  \item $C'_1$ is the set of clocks associated with $C_1$ as previously,
  and clocks in $X_1$ will be read on the synchronizations only,
  \item $E'_1
  \begin{array}[t]{ll}
    =&\{s\xrightarrow{z=d,\eps,r'\cup\{z\}}s'\mid \exists
    s\xrightarrow{d,e}s'\in E: e=(\ell(s)\xrightarrow{g,a,c(r')}\ell(s'))\}\\
    \cup&\{s\xrightarrow{\gamma,(a,\ell_2),c}s'\mid
    s \in S \land \gamma\equiv\gamma(R(s'))
    \land a \in\sync \land \exists\ell_1\xrightarrow{g,a,r}\ell_2\in E_1\}
  \end{array}$\\ 
  where 
  $\gamma(R(s'))$ is the clock constraint that describes the region of state $s'$,
  and
  $c$ still denotes the assignment of any clock $x'\in C'_1$
  with the value of its associated clock $c(x')=x\in C_1$ (written $x':=x$).
  \item $\forall s\in S$, $\inv'_1(s)\equiv z\leq d$ if $\exists s\xrightarrow{d,e}s'\in E$,
  and $\inv'_1(s)\equiv \true$ otherwise.
\end{itemize}


We can now prove the direct way of Theorem~\ref{thm:nsc} in this setting
where $A_1$ may have urgent synchronizations, and the most time consuming
local runs between two synchronizations exist.
First, let us recall some notations.
$\S_\mi{mod}=A'_1\parallel (A_{1,2}\x A_{2,\mi{mod}})$, with the same
$A'_1$ and $A_{2,\mi{mod}}$ as before,
$A'_2=A_{1,2}\x A'_{2,\mi{mod}}$ where $A'_{2,\mi{mod}}$ denotes
$A_{2,\mi{mod}}$ without location $\sad$,
and $\psi$ is such that for any
$((a,\ell_1),s_1)\in \sync'\times Q'_1$, $\psi(((a,\ell_1),s_1))=(a,s_1)$.
\proof[Proof of Theorem~\ref{thm:nsc}, 
when runs of maximal duration before synchronization exist]\hfill

\noindent We show that when $\noRest[A_1]{A_2}$ holds, $A_2$ does not need to read the clocks of
$A_1$, because then, the constructed $A'_1\parallel A'_2$ satisfies Definition~\ref{def:nsc},
\ie has no shared clocks and
  \begin{enumerate}
    \item $\psi(\TTS[Q_1']{A'_1}[A'_2])\wtb\TTS[Q_1]{A_1}[A_2]$ and
    \item $\psi(\TTS[Q'_1]{A'_1})\wtb\TTS[Q_1]{A_1}$ (this still holds because
    $A'_1$ has not changed)
    \item $\psi(\TTS[A'_1]{A'_2})\wtb\TTS[A_1]{A_2}$.
  \end{enumerate}

  First, we can prove that
  $\sad$ is reachable in $\S_\mi{mod}$ iff there is a restriction
  in $\TTS[A_1]{A_2}$, as we proved Lemma~\ref{lem:sad_equiv_read}.
  Indeed, what works when $A_{1,2}$ simulates any run of $A_1$ also
  works when $A_{1,2}$ simulates a fixed run of $A_1$.

  Then, we can prove that, if $\sad$ is not reachable (\ie if there is
  no restriction in $\TTS[A_1]{A_2}$),
  then $\psi(\TTS[A'_1]{A'_2})\wtb\TTS[A_1]{A_2}$.
  We use the same relation $\R$ as in the previous proof in~\ref{proof},
  that is, $\R$ is the relation such that for any reachable state $(S_1,s_2)$ of
  $\TTS[A_1]{A_2}$, and any reachable state $(S'_1,s'_2)$ of $\psi(\TTS[A'_1]{A'_2})$,
  \[(S_1,s_2)\R(S'_1,s'_2) \iffdef\left\{
  \begin{array}{l}
    s_2=(\ell_2, v_2)\text{ and }s'_2=((\ell_{1,2},\ell_2),v'_2)\text{ s.t.}\\
    \forall x\in X_2\setminus X_1, v_2(x)=v'_2(x)\\
    S_1=S'_1\\
  \end{array}
  \right.\]
  The proof of this bisimulation follows the same steps as the proof in~\ref{proof},
  except now we know that $A_{1,2}$ cannot force a synchronization by construction,
  and not by assuming that there is not urgent synchronization in $A_1$.

  Then, by Lemma~\ref{lem:nsc}, $\psi(\TTS[Q'_1]{A'_1}[A'_2])\wtb\TTS[Q_1]{A_1}[A_2]$
  also.\qed

\subsubsection*{Case 2: There is not always a Local Run of Maximal Duration after a Synchronization}
Now, we show how to adapt the previous construction when there are strict time
constraints and there is no path of maximal duration before an urgent
synchronization. For example, consider automaton $A_1$ of Fig.~\ref{fig:urgent2}
that has an urgent synchronization and such that there is no path of maximal duration
before this synchronization is taken: as previously, $b$ has to be performed
as late as possible, but because of the strict inequality $x<1$ on the edge
labeled by $b$, it is not possible to enforce this.

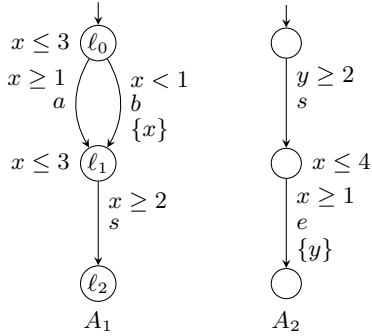
\begin{figure}[t]
  \centering
  \def\b{3.7}
\begin{tikzpicture}[node distance = 1.6cm, initial where = above]
  \node[state, initial] (q_0) [label=left:$x\leq3$] {$\ell_0$};
  \node[state] (q_1) [below of=q_0]
          [label=left:$x\leq3$] {$\ell_1$};
  \node[none] (a_1) at (0,-\b) {$A_1$};
  \node[state] (q_4) [below of=q_1] {$\ell_2$};
  
  \node[state, initial] (q_2) at (2.5,0) {};
  \node[state] (q_3) [below of=q_2] [label=right:$x\leq4$] {};
  \node[state] (q_5) [below of=q_3] {};
  \node[none] (a_2) at (2.5,-\b) {$A_2$};

  \path[->] (q_0) edge [bend right] node [swap, pos=0.45] {$x\geq1$}
                                    node [swap] {$a$} (q_1)
            (q_1) edge node [pos=0.25] {$x\geq2$}
                       node {$s$} (q_4)
            (q_0) edge [bend left] node [pos=0.45]{$x<1$}
                                    node {$b$}
                                    node [pos=0.55] {$\{x\}$} (q_1);

  \path[->] (q_2) edge node [pos=0.2]{$y\geq2$} node {$s$} (q_3)
            (q_3) edge node [pos=0.2] {$x\geq1$}
                                   node {$e$}
                                   node [pos=0.8] {$\{y\}$}(q_5);
\end{tikzpicture}
  \caption{$A_1$ has an urgent synchronization and there is no path with
  maximal duration before this synchronization.}\label{fig:urgent2}
\end{figure}

Here also, the construction relies on the region automaton and on the
computation of the supremum of the possible durations. Then the idea is again to
follow one of the paths with the best supremum duration. But there may not exist
any optimal timing to run this path and reach the supremum. Then we run it with
one possible timing and we wait in the last region, ignoring the invariant that
would force us to synchronize.
In our example, the supremum of the duration of the path
with $b$ is 4, and is greater than the supremum
of any other paths (the paths with $a$ have a maximal duration of 3).
Therefore, $b$ has to be performed while $x$ is in the region defined by
$0<x<1$.

Now, when $A_{1,2}$ reaches a state where it has to synchronize, if
$A'_1$ is not ready to synchronize (\ie $A'_1$ is not in the location
before the synchronization), then this means that $A'_1$ took
a more time consuming path (and not necessarily the same actions).
Then $A_{2,\mi{mod}}$ can stop using the values of the clocks of $A_{1,2}$ to
evaluate the truth value of its time constraints, 
and simply take their truth value according to the last region that makes
the invariant of the urgent synchronization true (\ie the region of its current
valuation), since it would still be in this region if it had been more time consuming.
Note that, if $\sad$ is not reachable, this means that, if $A_{1,2}$ had
performed a more time consuming run (for example the actual run followed by $A'_1$),
then $A_{2,\mi{mod}}$ would have been able to perform the same run.
Therefore, ``stopping'' the clocks in their current region has no
side effects.

In the construction, this results in new synchronization edges, performed
by $A_{1,2}$ and $A_{2,\mi{mod}}$, when $A_{1,2}$ has not been
slow enough (\ie when the invariant expires).
In our example, the synchronization labeled by $\mi{final\_region}_R$,
guarded by $x'=3$, notifies $A_{2,\mi{mod}}$ that $A_{1,2}$
is stuck in the final region $R$ (here $R$ corresponds to $\ell_1$ and $2
< x' < 3$) but that its clocks do not satisfy the constraint any more.
In this case, $A_{2,\mi{mod}}$ enters a duplicated version of itself, where
the guards over $X_1$ are no more evaluated according to the value of the
duplicated clocks $X'_1$, but simply replaced by their truth value according
to the final region.
In the example of Fig.~\ref{fig:result2}, the constraint $x' \geq 1$ that
appears on the arc from $\ell'$ to $\ell''$ is simply replaced by $\true$,
because the constraint is true in region $R$.
The duplicated versions can still reach location
$\sad$,
and the constraints on the edges leading to $\sad$ are also evaluated
according to the final region.

If a synchronization happens when $A_{2,\mi{mod}}$ is in one of its
duplicated versions,
then $A_{2,\mi{mod}}$ goes back to its initial version,
as depicted in Fig.~\ref{fig:result2}.


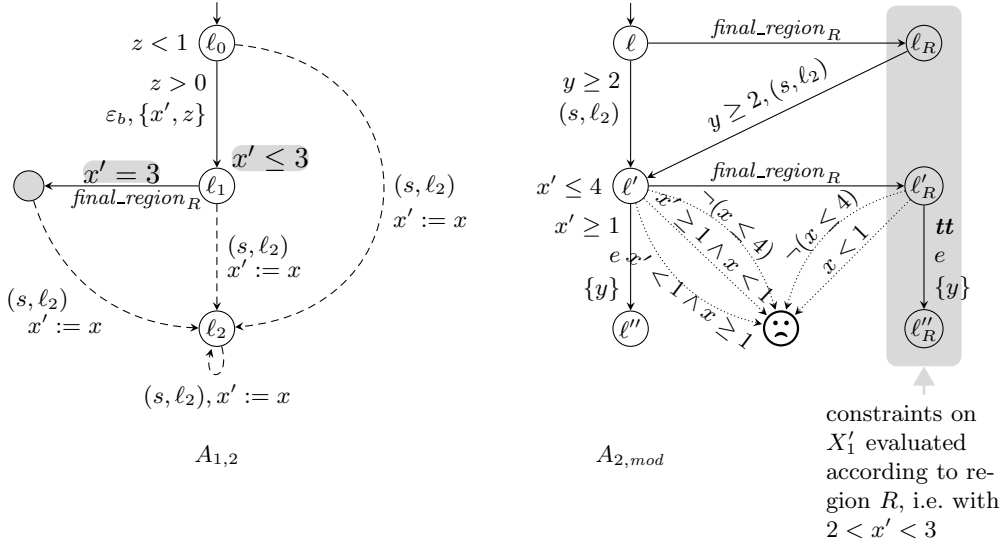
\begin{figure}[t]
  \centering
  \def\c{1.9}
  \def\b{5.5}
  \begin{tikzpicture}[node distance = \c cm, initial where = above]
  \fill[lightgray!60,rounded corners] (9.4-0.5,0+0.5) rectangle (9.4+0.5,-2*\c-0.5);
  \node[state, initial] (q_0) [label=left:$z<1$]{$\ell_0$};
  \node[state] (q_1) [below of=q_0,label={above right,inner sep=0.5pt}:{\colorbox{lightgray!60}{$x'\leq3$}}] {$\ell_1$};
  \node[state] (q_2) [below of=q_1] {$\ell_2$};
  \node[state,fill=lightgray!60] (q_5)  at (-2.5, -\c) {};
  \node[none] (a_1) at (0,-\b) {$A_{1,2}$};
  \path[->,swap] (q_0) edge node[pos=0.2] {$z>0$}
                       node {$\eps_b,\{x',z\}$}(q_1)
            (q_1) edge node[above, inner sep=1pt] {{\colorbox{lightgray!60}{$x'=3$}}}
        node[below, inner sep=1pt,pos=0.4] {\scriptsize$\mi{final\_region}_R$} (q_5)
                  ;
  \path[->,densely dashed](q_1) edge
                       node [pos=0.4] {$(s,\ell_2)$}
                       node [pos=0.6] {$x':=x$} (q_2)
            (q_0) edge [bend left=85, looseness=1.8]
                       node [pos=0.5] {$(s,\ell_2)$}
                       node [pos=0.53] {$x':=x$}(q_2)
            (q_2) edge [loop below] node {$(s,\ell_2),x':=x$}(q_2)
            (q_5) edge [swap,bend right=35]
                       node [pos=0.4] {$(s,\ell_2)$}
                       node [pos=0.6] {$x':=x$}(q_2);

  \node[state, initial] (q_2) at (5.5,0) {$\ell$};
  \node[state] (q_3) [below of=q_2] [label=left:$x'\leq4$] {$\ell'$};
  \node[state] (q_5) [below of=q_3] {$\ell''$};
  \node[sad] (q_6) at (7.5,-2*\c) {\huge\sad};

  \node[state] (q_7) [right of=q_2,xshift=2cm] {$\ell_R$};
  \node[state] (q_8) [right of=q_3,xshift=2cm] {$\ell'_R$};
  \node[state] (q_9) [right of=q_5,xshift=2cm] {$\ell''_R$};
  \node[none] (a_2) at (5.5,-\b) {$A_{2,\mi{mod}}$};
  \node[none,text width=2.6cm] (a_3) at (9.4,-\b-.2) {constraints on $X'_1$
    evaluated according to region $R$, i.e.\ with $2 < x' < 3$};
  \draw[->,lightgray!60,line width=1pt,>=triangle 60] (a_3)--(9.4,-2*\c-0.5);

  \path[->] (q_2) edge node [pos=0.2,swap] {$y\geq2$}
                       node [swap] {$(s,\ell_2)$} (q_3)
            (q_3) edge node [pos=0.2,swap] {$x'\geq1$}
                       node [swap]{$e$}
                       node [pos=0.8,swap] {$\{y\}$} (q_5)

        (q_2) edge node[above, inner sep=1pt] {\scriptsize$\mi{final\_region}_R$} (q_7)
        (q_3) edge node[above, inner sep=1pt] {\scriptsize$\mi{final\_region}_R$} (q_8)

            (q_7) edge node [sloped, above]{$y\geq2,(s,\ell_2)$} (q_3)
            (q_8) edge node [pos=0.2] {$\true$}
                       node {$e$}
                       node [pos=0.8] {$\{y\}$}(q_9)
            ;
  \path[->,densely dotted] (q_3) edge [bend left]
                       node [sloped, above, inner sep=-1pt] {$\neg(x\leq4)$}(q_6)
                  edge node [sloped, above, inner sep=1pt] {$x'\geq1\land x<1$}(q_6)
                  edge [bend right=30]
                       node [sloped, below, inner sep=-2pt,pos=0.6] {$x'<1\land x\geq1$}(q_6)
            (q_8) edge [bend right]
                       node [sloped, above, inner sep=-1pt] {$\neg(x\leq4)$}(q_6)
                  edge node [sloped, above, inner sep=1pt] {$x<1$}(q_6);
\end{tikzpicture}
  \caption[$A_{1,2}$ and $A_{2,\mi{mod}}$ for the NTA of Fig.~\ref{fig:urgent2}]
  {$A_{1,2}$ and $A_{2,\mi{mod}}$ for the NTA of Fig.~\ref{fig:urgent2}.}
  \label{fig:result2}
\end{figure}

In order to prove the soundness of the construction, one has to show that if
there is no restriction in $\TTS[A_1]{A_2}$ (\ie if $\sad$ is not reachable),
then $\psi(\TTS[A'_1]{A'_2})\wtb\TTS[A_1]{A_2}$. The bisimulation relation now
takes the new states into account as follows.
\[(S_1,s_2)\R(S'_1,s'_2) \iffdef\left\{
\begin{array}{l}
  s_2=(\ell_2, v_2)\text{ and }s'_2=((\ell_{1,2},\ell'_2),v'_2)\text{ s.t.}\\
  \text{\colorbox{lightgray!60}
    {$\ell_2=\ell'_2$ or $\ell'_2$ is one of the duplicated versions of $\ell_2$}}\\
  \forall x\in X_2\setminus X_1, v_2(x)=v'_2(x)\\
  S_1=S'_1\\
\end{array}
\right.\]

\section{Discussion and Extensions}\label{sec:shared_clocks_ext}

We have shown that in a distributed framework, when locality of actions and
synchronizations matter, NTA with shared clocks cannot be easily transformed
into NTA without shared clocks. The fact that the transformation is possible can
be characterized using the notion of contextual TTS which represent the
knowledge of one automaton about the other. Checking whether the transformation
is possible is \textsf{PSPACE}-complete.

In system design, our technique could help a designer to use shared clocks in an
abstract specification, and build automatically an implementable distributed
model without shared clocks.
Coming back to the example described in the introduction with several agents
performing together a distributed task according to a predefined schedule, this
would generate the mechanism for creating the local copies of the schedule.

A first point to notice is that, contrary to what happens when one considers the
sequential semantics, NTA with shared clocks are strictly more expressive if we
take distribution into account. This somehow justifies why shared clocks were
introduced: they are actually more than syntactic sugar.

Another interesting point that we want to recall here is the use of
transmitting information during synchronizations. In the end, when the
construction is possible, the only modification that is needed for $A_1$
is the renaming of the synchronizations, which codes this transmission of
information. On the other side, $A_2$ needs a much stronger modification in
order to handle the information transmitted by $A_1$.

Finally, it is noticeable that infinitely precise information is required in
general. This advocates the interest of updatable (N)TA used in an appropriate
way, and more generally gives a flavor of a class of NTA closer to
implementation.

\subsubsection*{Perspectives}
Our first perspective is to generalize our result to the symmetrical case where
$A_1$ also reads clocks from $A_2$. Then of course we can tackle general NTA
with more than two automata.

Notice that the set $\UR(s_1)$ used in the definition of contextual TTS is
always put in parallel with a state $s_2$. Therefore, it can be extended to
$\UR_{s_2}(s_1)$ that represents the set of states that $A_1$ can immediately
reach from $s_1$ while $A_2$ is in $s_2$. This means that the TTS of
$A_2$ in the context of $A_1$ can still be defined when
$A_1$ also reads clocks from $A_2$.
However, we do not know whether Theorem~\ref{thm:nsc} is still true with this
definition of contextual TTS, because most of the intermediate lemmas and
propositions to prove this theorem use $\TTS{A_1}$ that is not defined
when $A_1$ reads clocks from $A_2$.

Another line of research is to focus on transmission of information. The
goal would be to minimize the information transmitted during synchronizations,
and see for example where the limits of finite information lay.
Even when infinitely precise information is required to achieve the exact
semantics of the NTA, it would be interesting to study how this semantics can be
approximated using finitely precise information.

Finally, when shared clocks are necessary, one can discuss how to minimize their
number, or how to implement the model on a distributed architecture and how to
handle shared clocks with as few communications as possible.

\pagebreak
\bibliographystyle{alpha}
\bibliography{references}

\newcommand{\etalchar}[1]{$^{#1}$}
\begin{thebibliography}{BDM{\etalchar{+}}98}

\bibitem[ABG{\etalchar{+}}08]{ABGMN-concur08}
S.~Akshay, Benedikt Bollig, Paul Gastin, Madhavan Mukund, and K.~Narayan~Kumar.
\newblock Distributed timed automata with independently evolving clocks.
\newblock In {\em International Conference on Concurrency Theory (CONCUR)},
  volume 5201 of {\em LNCS}, pages 82--97, Toronto, Canada, 2008. Springer.

\bibitem[AD90]{AD90}
Rajeev Alur and David Dill.
\newblock Automata for modeling real-time systems.
\newblock In {\em Automata, Languages and Programming}, volume 443 of {\em
  LNCS}, pages 322--335. Springer, 1990.

\bibitem[AD94]{AD94}
Rajeev Alur and David Dill.
\newblock A theory of timed automata.
\newblock {\em Theoretical Computer Science}, 126(2):183--235, 1994.

\bibitem[BC12]{BC-concur12}
Sandie Balaguer and {\relax Th}omas Chatain.
\newblock Avoiding shared clocks in networks of timed automata.
\newblock In Maciej Koutny and Irek Ulidowski, editors, {\em {P}roceedings of
  the 23rd {I}nternational {C}onference on {C}oncurrency {T}heory
  ({CONCUR}'12)}, volume 7454 of {\em Lecture Notes in Computer Science},
  Newcastle, UK, September 2012. Springer.

\bibitem[BCH{\etalchar{+}}05]{BCHRL-FORMATS2005}
B{\'e}atrice B{\'e}rard, Franck Cassez, Serge Haddad, Didier Lime, and
  Olivier~H. Roux.
\newblock Comparison of the expressiveness of timed automata and time {P}etri
  nets.
\newblock In Paul Pettersson and Wang Yi, editors, {\em FORMATS}, volume 3829
  of {\em LNCS}, pages 211--225. Springer, 2005.

\bibitem[BCH12]{BCH-fmsd12}
Sandie Balaguer, {\relax Th}omas Chatain, and Stefan Haar.
\newblock A~concurrency-preserving translation from time {P}etri nets to
  networks of timed automata.
\newblock {\em Formal Methods in System Design}, 2012.

\bibitem[BDFP04]{upd}
Patricia Bouyer, Catherine Dufourd, Emmanuel Fleury, and Antoine Petit.
\newblock Updatable timed automata.
\newblock {\em Theoretical Computer Science}, 321(2-3):291--345, 2004.

\bibitem[BDKP91]{BestDKP91}
Eike Best, Raymond~R. Devillers, Astrid Kiehn, and Lucia Pomello.
\newblock Concurrent bisimulations in {P}etri nets.
\newblock {\em Acta Inf.}, 28(3):231--264, 1991.

\bibitem[BDL04]{uppaal}
Gerd Behrmann, Alexandre David, and Kim~Guldstrand Larsen.
\newblock A tutorial on {\sc uppaal}.
\newblock In Marco Bernardo and Flavio Corradini, editors, {\em Formal Methods
  for the Design of Real-Time Systems: 4th International School on Formal
  Methods for the Design of Computer, Communication, and Software Systems,
  SFM-RT 2004}, number 3185 in LNCS, pages 200--236. Springer--Verlag,
  September 2004.

\bibitem[BDM{\etalchar{+}}98]{kronos}
Marius Bozga, Conrado Daws, Oded Maler, Alfredo Olivero, Stavros Tripakis, and
  Sergio Yovine.
\newblock {\sc Kronos}: a model-checking tool for real-time systems.
\newblock In {\em CAV}, volume 1427 of {\em LNCS}, pages 546--550, 1998.

\bibitem[BDMP03]{Bouyer03}
Patricia Bouyer, Deepak D'Souza, P.~Madhusudan, and Antoine Petit.
\newblock Timed control with partial observability.
\newblock In Warren~A. Hunt, Jr and Fabio Somenzi, editors, {\em CAV 2003},
  volume 2725 of {\em LNCS}, pages 180--192. Springer, Heidelberg, 2003.

\bibitem[BHR06]{BHR-atva06}
Patricia Bouyer, Serge Haddad, and Pierre-Alain Reynier.
\newblock Timed unfoldings for networks of timed automata.
\newblock In Susanne Graf and Wenhui Zhang, editors, {\em {P}roceedings of the
  4th {I}nternational {S}ymposium on {A}utomated {T}echnology for
  {V}erification and {A}nalysis ({ATVA}'06)}, volume 4218 of {\em LNCS}, pages
  292--306, Beijing, China, October 2006. Springer.

\bibitem[BJLY98]{Bengtsson98partialorder}
Johan Bengtsson, Bengt Jonsson, Johan Lilius, and Wang Yi.
\newblock Partial order reductions for timed systems.
\newblock In {\em CONCUR}, volume 1466 of {\em LNCS}, pages 485--500. Springer,
  1998.

\bibitem[BR08]{BoyerR08}
Marc Boyer and Olivier~H. Roux.
\newblock On the compared expressiveness of arc, place and transition time
  {P}etri nets.
\newblock {\em Fundamenta Informaticae}, 88(3):225--249, 2008.

\bibitem[CCJ06]{Cassez-Ch-Jard_ATVA06}
Franck Cassez, {\relax Th}omas Chatain, and Claude Jard.
\newblock Symbolic unfoldings for networks of timed automata.
\newblock In {\em ATVA}, volume 4218 of {\em LNCS}, pages 307--321. Springer,
  2006.

\bibitem[CGL93]{Epsilon93}
Karlis Cerans, Jens~Chr. Godskesen, and Kim~Guldstrand Larsen.
\newblock Timed modal specification - theory and tools.
\newblock In Costas Courcoubetis, editor, {\em CAV}, volume 697 of {\em LNCS},
  pages 253--267. Springer, 1993.

\bibitem[CR06]{Cassez-JSS06}
Franck Cassez and Olivier~H. Roux.
\newblock Structural translation from time {P}etri nets to timed automata.
\newblock {\em Journal of Systems and Software}, 2006.

\bibitem[Dim09]{Dima09}
C\u{a}t\u{a}lin Dima.
\newblock Positive and negative results on the decidability of the
  model-checking problem for an epistemic extension of timed {CTL}.
\newblock In {\em TIME}, pages 29--36. IEEE Computer Society, 2009.

\bibitem[DL07]{Dima}
C\u{a}t\u{a}lin Dima and Ruggero Lanotte.
\newblock Distributed time-asynchronous automata.
\newblock In {\em ICTAC}, pages 185--200. Springer-Verlag, 2007.

\bibitem[DLLN09]{David09}
Alexandre David, Kim~G. Larsen, Shuhao Li, and Brian Nielsen.
\newblock Timed testing under partial observability.
\newblock In {\em ICST}, pages 61--70. IEEE Computer Society, 2009.

\bibitem[HFMV95]{ReasoningAboutKnowledge}
Joseph~Y. Halpern, Ronald Fagin, Yoram Moses, and Moshe~Y. Vardi.
\newblock {\em Reasoning About Knowledge}.
\newblock MIT Press, 1995.

\bibitem[LNZ05]{LugiezNZ05}
Denis Lugiez, Peter Niebert, and Sarah Zennou.
\newblock A partial order semantics approach to the clock explosion problem of
  timed automata.
\newblock {\em Theoretical Computer Science}, 345(1):27--59, 2005.

\bibitem[LPW07]{LPW07}
Alessio Lomuscio, Wojciech Penczek, and Bozena Wozna.
\newblock Bounded model checking for knowledge and real time.
\newblock {\em Artif. Intell.}, 171(16-17):1011--1038, 2007.

\bibitem[Mer74]{Merlin}
Philip~Meir Merlin.
\newblock {\em A study of the recoverability of computing systems}.
\newblock PhD thesis, University of California, Irvine, 1974.

\bibitem[Min99]{Minea99}
Marius Minea.
\newblock Partial order reduction for model checking of timed automata.
\newblock In {\em CONCUR}, volume 1664 of {\em LNCS}, pages 431--446. Springer,
  1999.

\bibitem[Rei84]{Reif}
John Reif.
\newblock The complexity of two-player games of incomplete information.
\newblock {\em Jour. Computer and Systems Sciences}, 29:274--301, 1984.

\bibitem[Srb08]{Srba08}
Ji\v{r}\'{\i} Srba.
\newblock Comparing the expressiveness of timed automata and timed extensions
  of {P}etri nets.
\newblock In {\em FORMATS}, volume 5215 of {\em LNCS}, pages 15--32. Springer,
  2008.

\bibitem[vGG01]{GlabbeekG01}
Rob~J. van Glabbeek and Ursula Goltz.
\newblock Refinement of actions and equivalence notions for concurrent systems.
\newblock {\em Acta Inf.}, 37(4/5):229--327, 2001.

\bibitem[WL04]{WoznaL04}
Bozena Wozna and Alessio Lomuscio.
\newblock A logic for knowledge, correctness, and real time.
\newblock In {\em CLIMA}, volume 3487 of {\em LNCS}, pages 1--15. Springer,
  2004.

\end{thebibliography}

\end{document}